\documentclass[useAMS,usenatbib]{mn2e}
\usepackage{graphicx,multirow,url,mathrsfs,amssymb,amsmath,relsize,psfrag}
\def\n7{NGC\,7314}
\def\fa{Fe K$\alpha$~}
\def\rg{$r_{\rm g}$}
\def\ms{M$_{\rm \sun}$}
\newcommand{\eqb}{\begin{eqnarray}}
\newcommand{\eqe}{\end{eqnarray}}

\newcommand{\re}{\operatorname{Re}}
\newcommand{\im}{\operatorname{Im}}
\def\aj{AJ}%
\def\apj{ApJ}%
\def\apjl{ApJ}%
\def\apjs{ApJS}%
\def\apss{Ap\&SS}%
\def\aap{A\&A}%
\def\mnras{MNRAS}%
\def\pre{Phys.~Rev.~E}%
\def\nat{Nature}%
\DeclareMathOperator{\dex}{dex}
\DeclareMathAlphabet{\mathpzc}{OT1}{pzc}{m}{it}
\title[X-ray variability studies of \n7]{Extensive X-ray variability studies of \n7 using long \textit{XMM-Newton} observations\thanks{Based on observations obtained with \textit{XMM-Newton}, an ESA science mission with instruments and contributions directly funded by ESA Member States and NASA.}}
\author[D.~Emmanoulopoulos et al.]{D.~Emmanoulopoulos,$^{1}$\thanks{E-mail: D.Emmanoulopoulos@soton.ac.uk} I.~M.~M\textsuperscript{c}Hardy,$^{1}$ S.~Vaughan,$^{2}$ and I.~E.~Papadakis$^{3,4}$
\\
$^{1}$Physics and Astronomy, University of Southampton, Southampton, SO17 1BJ, UK\\
$^{2}$University of Leicester, X-Ray and Observational Astronomy Group, Department of Physics and Astronomy, Leicester, LE1 7RH, UK\\
$^{3}$Department of Physics and Institute of Theoretical and Computational Physics, University of Crete, 71003 Heraklion, Greece\\
$^{4}$IESL, Foundation for Research and Technology, 71110 Heraklion, Greece
}
\begin{document}

\date{Accepted 2016 May 10. Received 2016 May 5; in original form 2016 February 23}
\pagerange{\pageref{firstpage}--\pageref{lastpage}} \pubyear{2002}
\maketitle

\label{firstpage}
\begin{abstract}
We present a detailed X-ray variability study of the low mass Active Galactic Nuclei (AGN) \n7 using the two newly obtained \textit{XMM-Newton} observations (140 and 130 ks), together with two archival data sets of shorter duration (45 and 84 ks). The relationship between the X-ray variability characteristics and other physical source properties (such as the black hole mass) are still relatively poorly defined, especially for low-mass AGN. We perform a new, fully analytical, power spectral density (PSD) model analysis method, which will be described in detail in a forthcoming paper, that takes into consideration the spectral distortions, caused by red-noise leak. We find that the PSD in the $0.5-10$ keV energy range, can be represented by a bending power-law with a bend around $6.7\times10^{-5}$ Hz, having a slope of $0.51$ and $1.99$ below and above the bend, respectively. Adding our bend time-scale estimate, to an already published ensemble of estimates from several AGN, supports the idea that the bend time-scale depends linearly only on the black hole mass and not on the bolometric luminosity. Moreover, we find that as the energy range increases, the PSD normalization increases and there is a hint that simultaneously the high frequency slope becomes steeper. Finally, the X-ray time-lag spectrum of \n7 shows some very weak signatures of relativistic reflection, and the energy resolved time-lag spectrum, for frequencies around $3\times10^{-4}$ Hz, shows no signatures of X-ray reverberation. We show that the previous claim about ks time-delays in this source, is simply an artefact induced by the minuscule number of points entering during the time-lag estimation in the low frequency part of the time-lag spectrum (i.e.\ below $10^{-4}$ Hz).  

\end{abstract}


\section{Introduction}
\label{sect:intro}
X-ray variability studies of Active Galactic Nuclei (AGN) have been proven an excellent diagnostic tool of probing and disclosing their physical properties. The observed X-ray flux variations are thought to originate from the innermost region of an accretion-flow taking place around a super-massive black hole (BH), having a mass range between $10^5$ and $10^9$ \ms. Since the dynamics of this region are tuned by the BH mass, the corresponding size-scales, and thus time-scales, should depend on the BH mass via some sort of scaling relations. Moreover, since the accretion physics laws appear to be the same for all the classes of accretion objects, irrespective of their BH mass, these scaling laws should extend to much smaller `stellar-size' BH masses, such as those of the black-hole X-ray binary (BH-XRB) systems, having masses around $1-20$ \ms.\par
Currently, such an AGN-XRB connection is seen in the Fourier domain \citep{uttley02,markowitz03,vaughan03b,mchardy06}. Both classes of objects exhibit random X-ray flux variations whose amplitude, at a given Fourier frequency $f$, is described by Power Spectral Density (PSD) functions of a power-law form, $\mathscr{P}(f)\propto f^{-2}$, which break/bend below some characteristic frequency, $f_{\rm b}$, to flatter values. Exactly this characteristic frequency, $f_{\rm b}$, scales approximately inversely proportional to the BH mass. Notably, this scaling law seems to be followed only from those XRBs in the `soft-state', since those in the `hard-states' are characterised by more complex shapes with multiple bends \citep{nowak00,grinberg14}.\par
Moreover, this AGN-XRB connection is further supported by the X-ray reverberation studies. The detection of Fourier resolved time delays between the soft, $0.3-1$ keV, and the hard, $1-4$ keV, X-ray light curves, is attributed to the light-travel distance between the X-ray emitting corona (hard X-ray component) and the reflected emission (delayed, soft X-ray component) from the inner parts of the accretion disc, close to the BH, where strong gravity effects play a major role \citep{fabian09,emmanoulopoulos11b,zoghbi11,cackett13,fabian13,alston14}. \citet{demarco13} showed from an observational perspective the existence of a scaling relation between the most-negative time-lag measurement and the BH mass in a sample of 15 AGN. Then, \citet{emmanoulopoulos14}, using a fully general relativistic approach, were able to model the time-lag spectra of 12 AGN and derive/constrain physical source parameters, principally the BH mass and the height of the X-ray corona and then the BH spin parameter and viewing angle of 
the systems. Similar results on individual sources, focusing mainly in the height of the X-ray corona and the viewing angle, have been also derived by \citet{wilkins13,cackett14}.\par
Currently, with respect to galactic sources, such as XRBs, the detection of X-ray time delays is very challenging. Assuming that all the time-scales are scaling according to the BH mass i.e.\ $t_{\rm g}=r_{\rm g}/c$ where $t_{\rm g}$ is the light crossing time and \rg is the gravitational radius, for an object with $M=$10 $M_\odot$ $t_{\rm g}$ is around 50 $\mu$s, in contrast to and AGN with mass $M=2\times10^6$ $M_\odot$ for which $t_{\rm g}$=10 s. Thus, direct studies of X-ray time delays in the case of XRBs are prohibitively difficult to be conducted with the current instrumentation in the $\mu$s time-regime. Nevertheless, in the case of BHXRB GX-339 ms time-delys have been observed in a hard state by \citep{uttley11} using the covariance spectrum. Moreover, for X-ray lags have been reported for the neutron star X-ray binaries 4U\,1608-52 and 4U\,1636-53 are observed at the frequencies of the kHz  QPOs \citep{mcclintock13}. Interestingly, by joining this `galactic' time-scale sample, with the one derived 
by the AGN \citet{demarco13b} shows that the observed time-delays appear to scale with the BH mass \citep[figure~9 in][]{demarco13b}. In the same figure one can notice the great gap in the BH mass parameter space between galactic and extragalactic sources (almost 6 orders of magnitude).\par
Consequently, it is of great importance to bridge the gap, in the BH mass parameter space, between extragalactic and galactic accreting objects. Thus, the study of extragalactic low-mass BH mass objects, exhibiting a strong variable X-ray flux behaviour, provide us with a reliable test-bed to address the validity of such scaling relations along the lower BH mass-end of the parameter space which is currently a completely unexplored region.\par
Along these lines, \n7 \citep[z=0.004760,][]{mathewson96}, is an ideal source for such timing studies. \n7 is a spiral galaxy, morphologically classified as SAB(rs)bc, which hosts an AGN. The optical spectrum of \n7 is dominated by narrow emission lines but a weak broad wing to H${\alpha}$ has been noted by \citet{filippenko84} and \citet{hughes03}, leading to a tentative classification as a Seyfert type 1.9. However an O\,{\sc i} Bowen fluorescence line (at $\lambda=8446$ \AA) was noted by \citet{morris85}, leading to a classification as a Seyfert type 1 galaxy. The central stellar velocity dispersion is 60 km s$^{-1}$ \citep{gu06}, yielding a BH-mass estimate of $(0.87\pm0.45)\times10^6$ \ms \citep{mchardy13}. Finally, in the literature, there are three bolometric luminosity estimates of \n7: 6.46, 2.63 and 9.55 $(\times10^{43}$ erg s$^{-1})$ \citep{elvis94,marconi04,vasudevan07} and thus in our analysis we are using their mean value $(6.21\pm3.47)\times10^{43}$ erg s$^{-1}$. Finally, from the very early years of X-ray astronomy, \n7 showed a great deal of flux variations spanning minutely \citep[\textit{EXOSAT} data,][]{turner87} up to hourly flux variations \citep[\textit{ASCA} data,][]{yaqoob96}. There is observational evidence that the source is seen through a warm absorber which is situated within the clumpy torus \citep{ebrero11}. In this case variations in the neutral column density are explained in terms of a cloud of neutral gas crossing our line of sight, which graze the edge of the torus.\par
In this paper we present a thorough X-ray variability analysis of \n7 in the Fourier domain, using the PSD method and the time-lag spectra. Initially, in Sect.~\ref{sect:obs_datRed} we present the X-ray data-reduction procedures for the observations obtained by the European photon imaging camera (EPIC), consisting of the pn-charge coupled device (pn-CCD) and the two metal oxide semi-conductor (MOS)-CCDs. In the next section we estimate the fractional room mean square (rms) variability amplitude as a function of X-ray energy and we study the rms-flux relation. Then, in Sect.~\ref{sect:pds_model} we describe briefly the new PSD modelling and fitting procedure, that we use for this work, which will be extensively described in a forthcoming paper. Moreover, in the same section we present the PSD fitting results in several X-ray energy bands. Following that, in Sect.~\ref{sect:scale_relations} we revisit the PSD scaling relation, using our PSD bend-time scale estimates, and the values from an already published 
ensemble of 15 AGN. In Sect.~\ref{sect:tlspec} we present a detailed X-ray time-lag analysis for \n7, the corresponding lamp-post general relativistic model and the energy resolved time-lag spectrum. Finally, a discussion together with a summary of our results can be found in the last section (Sect.~\ref{sect:summ_disc}). Throughout the paper the error estimates for the various physical parameters correspond to the 68.3 per cent confidence intervals unless otherwise stated. Similarly, the error bars of the plot points in all the figures indicate the 68.3 per cent confidence intervals.

\section{OBSERVATIONS AND DATA-REDUCTION}
\label{sect:obs_datRed}
\subsection{\textit{XMM-Newton} observations}
\label{ssect:obs}
\n7 was observed by \textit{XMM-Newton} twice during 2013. The first observation (Obs. ID: 0725200101) started on 2013 May 17, 02:59:08 (UTC) and lasted for 140473 (s). The second observation (Obs. ID: 0725200301) started on 2013 November 28, 15:36:13 (UTC) with an on-source duration time of 131013 s. In both observations the pn camera was operated in Prime Small Window mode (time resolution: 6 ms) and the two MOS cameras in Prime Full Window mode (time resolution: 2.6 s). Medium-thickness aluminised optical blocking filters (80 nm Al and 160 nm polyimide) were used for all EPIC cameras to reduce the contamination of the X-rays from infrared, visible, and ultra-violet light.\par
Prior to these two recent long observations, \n7 was observed twice. The first observation (Obs. ID: 0111790101) was obtained on the 2001 May 02, 09:56:21 (UTC) and lasted for 44663 s. The EPIC camera-modes as well as the filters were the same with those used for the 2013 observations. Finally, \n7 happens to be in the same field of view with one of the most distant massive galaxy cluster XMMU\,J2235.3-2557 (z=1.4). This galaxy cluster was observed on the 2006 May 03, 12:41:07 (UTC) with \textit{XMM-Newton} \citep[Obs. ID: 0311190101,][]{baldi12} for 83920 s, and \n7 was located 7.7 arcmin south-west from the centre of the field of view. Both MOS and pn cameras were in Prime Full Window mode (the time resolution for the pn is 73.4 ms) and used the thin filter set-up (40 nm Al and 160 nm polyimide) whose optical blocking is expected to be about 100 times less efficient than the medium-thick filter. The galaxy cluster lies in the centre of the field of view and it is several orders of magnitude fainter than \n7, thus it does not contaminate the flux levels of our source.  

\subsection{\textit{XMM-Newton} data reduction}
\label{ssect:datRed}
The EPIC raw-data are reduced with the \textit{XMM-Newton} {\sc scientific analysis system} ({\sc sas}) \citep{gabriel04} version 14.0.0. After reprocessing the pn and the two MOS data-sets with the \textit{epchain} and the \textit{emchain} {\sc sas}-tools respectively, we perform a thorough check for pile-up using the task \textit{epatplot}. Neither the pn nor the two MOS data sets appear to suffer from pile-up in any observation, therefore count-rates for the source and the background are extracted from circular regions that are given in Table~\ref{tab:datRed}. We also verify by visual inspection that all the resulting light curves from all the observations are not affected by the pile-up effect.\par 
For the extraction of the pn light curves, at a given energy band, we allow events that are detected up to double pixel-pattern on the CCD (PATTERN$\leq$4) and we exclude all the events that are at the edge of a CCD and at the edge to a bad pixel i.e.\ FLAG$=$0. For the MOS light curves we select events that are detected up to quadruple pixel-pattern on the CCDs (PATTERN$\leq$12) and we restrict to those events having also FLAG$=$0. The corrected background-subtracted light curves of the source are produced using the {\sc sas}-tool \textit{epiclccorr} in time-bins of 20 s. Note, that this tool takes correctly into account the vignetting effects that may affect the 2006 off-centre observation of the source (Obs. ID: 0311190101).\par
\begin{figure}
\includegraphics[width=3.2in]{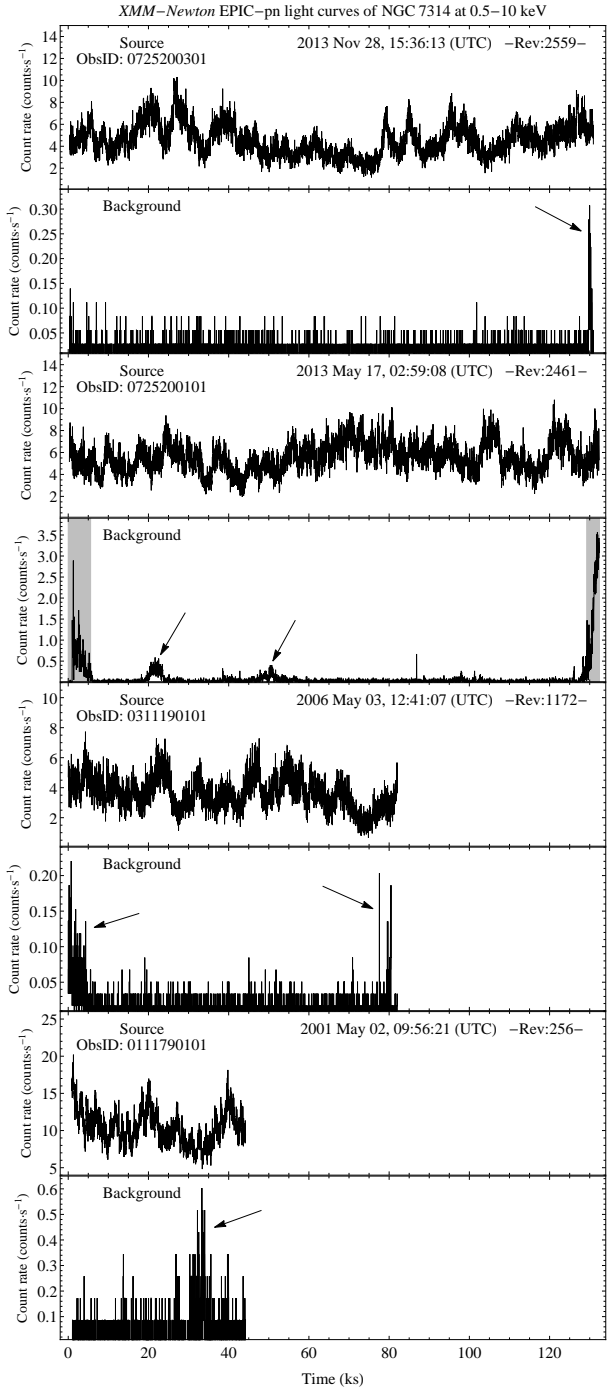}
\caption{Source and background EPIC-pn light curves in the $0.5-10$ keV energy range. The arrows indicate observing periods of background activity, reaching levels of the order of 4 to 5 per cent of the source's count-rate. Due to their small contribution-ratio these periods are kept in our analysis. The grey regions in the fourth panel, indicate the high-background levels, occurring during the observation 0725200101, for the first 5.6 ks and the last 3.2 ks. These periods are excluded from our analysis.}
\label{fig:lcs}
\end{figure}

Using the EPIC-pn background light curves, in the $0.5-10$ keV energy range (Fig.~\ref{fig:lcs}), we can monitor the background activity during the course of the observations. For the first observation, 0725200301, there is no need to subtract any parts from the light curves, yielding a total of 129.86 ks of useful data. The apparent increase towards the end of the observation (indicated by the arrow in Fig.~\ref{fig:lcs}, second panel) is only 5 per cent of the source's count rate (source$/$background$=6/0.3$) and thus, it is properly taken into account during the background subtraction. The second observation, 0725200101, has two very prominent increased background levels, (Fig.~\ref{fig:lcs}, fourth panel, grey regions) during the first 5.6 ks and the last 3.2 ks of the observation that reach up to 50 per sent of the source count-rate. Thus, data from those periods are disregarded from our analysis and thus we end up with 123.66 ks of continuous observations. There are also two background `bumps' at around 21 and 50 ks, respectively, indicated by the arrows, but these are only 5 per cent of the source's count-rate and thus we keep these segments in our analysis. The third and the fourth observations, 0311190101 and 0111790101 respectively, are left intact since the apparent increase activity is of the order of 4 per cent of the source's count rate. Thus, from these two archival observations we have additionally for our analysis a continuous stream of data lasting 82.04 and 43.08 ks, respectively. Finally, for a given energy band, we merge the light curves coming from the three EPIC instruments (the pn and the two MOS detectors) into a combined light curve, achieving in this way the maximum signal to noise ratio.\par
It is known that the MOS-to-pn results are broadly very similar with MOS excesses over pn at a maximum of 8 per cent \citep{read14}. Particularly for our four data sets, we performed a consistency check by estimating the ratio between the MOS and the pn light curves in the 0.5-10 keV energy range. For each observation we fitted the ratios with a linear model and we found that more than 99 per cent of the ratio points lie within 3 standard deviations ($\sigma$) from the best-fitting model which has on average a value of around 0.31 (the best-fitting slope is always practically zero -of the order of $10^{-8}$-). Following the same procedure the agreement between the two MOS detectors is much better since for all observations more than 99.98 per cent of all observations lien within 3$\sigma$ from the best-fitting linear model.

\begin{table}
\begin{minipage}{85mm}
\caption{Source and background regions for the \textit{XMM-Newton} EPIC light curves. The first column, (1), gives the observation ID, the second column, (2), the EPIC instrument, the third column, (3), the circular radius centred around the source, and the fourth column, (4), the circular background extraction region in the form of radius and centre-position.}
\label{tab:datRed}
\begin{tabular}{@{}cccc}
\hline
(1) & (2) & (3) & (4) \\
Obs. ID &     EPIC   & Source        & Background             \\
        & instrument & radius        &  radius and centre     \\
        &           &  (arcsec)     &  (arcsec, physical units\footnote{These are the projected sky coordinates which are denoted in the FITS files by X and Y.})   \\
\hline\hline
           & pn		&  36.60	&  49.00   (28175, 23457) \\
0725200301 & MOS-1	&  50.60	&  113.40  (30120, 21771) \\
           & MOS-2	&  50.60	&  113.40  (30120, 21771) \\ \hline
           & pn		&  39.15	&  53.50   (23786, 28549) \\
0725200101 & MOS-1	&  52.80	&  109.00  (26222, 19830) \\
           & MOS-2	&  52.80	&  109.00  (26222, 19830) \\ \hline
           & pn		&  37.00	&  63.55   (13590, 16102) \\
0311190101 & MOS-1	&  27.75	&  189.00  (21608, 12119) \\
           & MOS-2	&  22.50	&  154.15  (24830, 12442) \\ \hline         
           & pn		&  41.10	&  31.35   (23983, 28181) \\ 
0111790101 & MOS-1	&  48.35	&  161.25  (34178, 18398) \\
           & MOS-2	&  45.90	&  150.50  (21598, 36387) \\ \hline
\end{tabular}
\end{minipage}
\end{table}

\section{THE RMS VARIABILITY AMPLITUDE}
\label{sect:fvar}
A very simple way to characterise the source's intrinsic variability amplitude is the fractional rms variability amplitude, $F_{\rm var}$ \citep{vaughan03}, which is defined as the square root of the noise-subtracted variance (i.e.\ excess variance, $\sigma_{\rm XS}^2$), divided by the mean count rate of the light curve. We estimate $F_{\rm var}$ from each observation (Sect.\ref{ssect:obs}), at a given energy band, and then we derive an average from all four values, which are 0.223, 0.246, 0.211 and 0.193 for $0.5-2$, $2-4$, $4-5$ and $5-10$ keV, respectively\footnote{The weighted mean errors that we derive for each energy band from the four observations are of the order of $(1-3)\times10^{-5}$.}. Thus, as the energy increases $F_{\rm var}$ decreases above 4 keV, a behaviour which is seen before in several AGN \citep[e.g.][]{green93,nandra97,mchardy04}.\par
\begin{figure}
\includegraphics[width=3.2in]{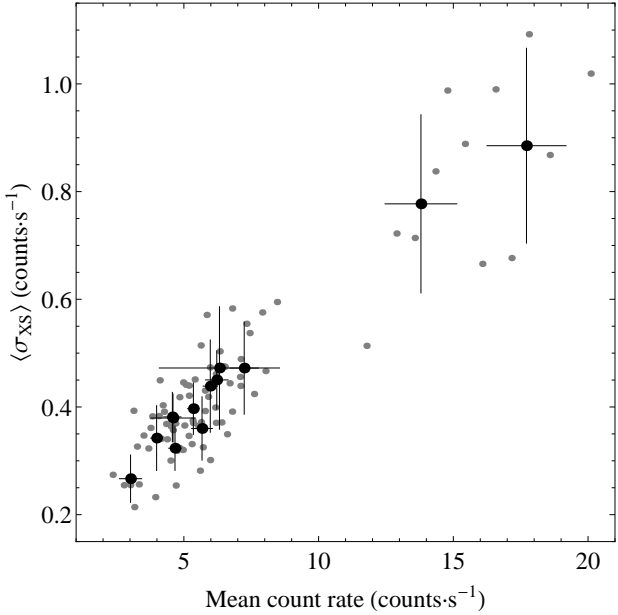}
\caption{The rms-flux relation for \n7 in the $0.5-10$ keV energy band. The grey points correspond to all the individual segments (lasting 240 s) and the black points to the corresponding average values of 10 consecutive segments.}
\label{fig:fluxRMS}
\end{figure}

Moreover, in order to characterise the relationship between the rms variability and the mean flux, we estimate the rms-flux relation in the $0.5-10$ keV energy band. The rms-flux relation requires that the amplitude in small time-scales is modulated by longer time-scale trends in data \citep{uttley05a}. We partition each light curve into segments of 240 s and we estimate for each segment the excess variance, $\sigma_{\rm XS}^2$ and the mean flux value. In Fig.~\ref{fig:fluxRMS} we show the rms amplitude $\sigma_{\rm XS}$ as a function of mean flux for all the individual segments of all four observations (grey points) and for the average values of 10 consecutive segments (black points). It is clear that the linear rms-flux relation holds for \n7 as in the case of several other AGN \citep[e.g.][]{vaughan03b,gaskell04,markowitz07,vaughan11}.

\section{PSD MODELLING: A FULLY ANALYTICAL APPROACH}
\label{sect:pds_model}
The PSD, $\mathscr{P}(f;\bmath{\gamma})$, with parameters $\bmath{\gamma}$, expressing slopes, breaks/bends, etc., is a continuous function that defines the amount of variability power at a given frequency, $f$. A statistical `natural' estimator of the PSD is the periodogram, $P(f_j)$, which is a discrete function, defined at the Fourier frequencies, $f_j$. Consider a light curve $x(t)$ with a mean value $\mu$, consisting of $N$ equidistant observations, with a sampling period $t_{\rm bin}$. The periodogram is the modulus square of the discrete Fourier transform estimated at $N/2$ evenly spaced frequencies $f_j=j/(Nt_{\rm bin})$ with $j=1,\ldots, N/2$ or $(N-1)/2$ for even or odd $N$, respectively. The exact definitions and properties of $\mathscr{P}(f)$ and $P(f_j)$ can be found in \citet{priestley81}. Throughout this work we apply the fractional rms squared normalization: $(2 t_{\rm bin})/(N \mu^2)$, for which the constant Poisson noise level is equal to $2(\mu+\left<\mu_{\rm bkg}\right>)/\mu^2$ and
$\left<\mu_{\rm bkg}\right>$ is the mean background count rate.\par
It is very well known that the PSD modelling procedure, through the periodogram estimation, can occasionally become a very complicated task when windowing effects are introduced in the Fourier domain. In the case of a non-white noise PSD (as in the case of AGN light curves) the periodogram estimates tend to be biased due to `red-noise leak' \citep[the transfer of variability power from the low to high-frequencies due to the finite length of observations;][]{deeter82,deeter84} and `aliasing effects' \citep[fold-back of variability power from high-frequencies to lower frequencies due to the finite time resolution;][]{kirchner05}. Additionally, missing data from a uniformly sampled data set (i.e.\ the data set consists of $N$ consecutive time bins of $t_{\rm bin}$ duration and part of these bins do not contain observations) perplexes even more the classical Fourier analysis, introducing statistical dependencies to the various periodogram estimates. As noted by \citet{babu10}, in the case of non-uniform sampling 
(irregular sampling) periodogram analysis can not be performed due to the lack of a well defined Nyquist frequency, however other methods based on interpolation, slotted resampling etc. are applicable.\par
Interestingly, one of the the most important  point for the case of all spectral distortions \citep[see for a discussion][]{deeming75}, is that they are introduced to the Fourier analysis \textit{only} via the sampling properties of the data set (i.e.\ length, bin-size and regularity). These properties are fully depicted by the window function, which can be described by a boxcar $w_{t}=1$ or $0$ (with or without observation, respectively). In the ideal case of the absence of spectral distortions, the periodogram estimates, at a given Fourier frequency, are distributed asymptotically around the underlying PSD as $\chi^2$ distribution with 2 d.o.f., $\chi^2_2$. However, if the periodogram estimates are affected by spectral distortions, then we get deviations from the expected $\chi^2_2$ distribution.\par
In this paper we implement a completely new fully analytical approach to model and fit the underlying PSD. The overall procedure is described in detailed in Emmanoulopoulos et al. 2016 (finished) (EMM16). The main idea can be briefly described as follows: since all the spectral distortions are entirely tuned by the window function of the observed data set, we convolve the given PSD model, $\mathscr{P}(f;\bmath{\gamma})$, with the observed window function. In this way, we introduce, in a mathematically precise way, all the spectral pathologies to the PSD model, for all Fourier frequencies, $f_j$. This is achieved by convolving the model autocovariance function, which is directly related with the model PSD via the inverse Fourier transform, with the observed window function. The resulting window-affected PSD model (distorted PSD model, $\mathscr{P}_{\rm d}(f_j;\bmath{\gamma})$), corresponds to the mean value of the (distorted) periodogram estimates, at a given $f_j$, and is given by

\hspace*{-1.9em}\parbox{1.08\linewidth}{
\eqb
\mathscr{P}_{\rm d}(f;\bmath{\gamma})&=&\frac{4t_{\rm bin}}{\mu^2N}\left\{\rule{0cm}{0.8cm}\right.\!2\pi N\int_{0}^{+\infty}\mathscr{P}(f;\bmath{\gamma})df+\nonumber\\
&&4\pi\mathlarger{\mathlarger{\mathlarger{\sum}}}_{s=1}^{N-1}\left\{\rule{0cm}{0.7cm}\right.\left(\int_{0}^{+\infty}\mathscr{P}(f;\bmath{\gamma})\cos\left[2\pi f t_s\right]df\right)\times\nonumber\\
&&\sum_{k=1}^{N-|s|}w_kw_{k+|s|}\cos\left[2\pi f_j t_s\right]\!\left.\rule{0cm}{0.7cm}\right\}\!\left.\rule{0cm}{0.8cm}\right\} 
\label{eq:final_aver_peri}
\eqe
}

Note that the two integrals in the above equation are the inverse Fourier transform of the autocovariance function at $t_s=0$ and $t_s\neq0$, and $t_s=st_{\rm bin}$ $(s=1,\ldots,N$). The (distorted) periodogram estimates, follow for each $f_j$ a gamma distribution, $\mathscr{G}\left[\nu_j/2, \mathscr{P}(f_j;\bmath{\gamma})\right]$, (based on simulations discussed in EMM16) around the mean $\mathscr{P}_{\rm d}(f_j;\bmath{\gamma})$, for which we do not know the d.o.f., $\nu_j$, (appearing through the shape parameter, $\nu_j/2$). The distorted PSD model is estimated via equation~\ref{eq:final_aver_peri} and this can be used further to derive precisely $\nu_j$, at each Fourier frequency, $f_j$. Both the shape and scale parameter of each gamma distribution at each $f_j$, are related with the mean, $\mathscr{P}_{\rm d}(f_j;\bmath{\gamma})$, via the following relation

\eqb
\nu_j=2\frac{\mathscr{P}_{\rm d}(f_j;\bmath{\gamma})}{\mathscr{P}(f_j;\bmath{\gamma})}
\label{eqe:dofDistorted}
\eqe

Finally, during the maximum likelihood approach, the final joint probability consists of an ensemble of gamma distributions with different shape parameters, each one carrying all the spectral distortions and dependencies, at a given Fourier frequency, $f_j$.\par
The method is fully analytical with respect to the derivation of the distorted PSD model (equation~\ref{eq:final_aver_peri}) and the corresponding gamma distribution of the distorted periodogram estimates, since it does not involve any assumptions and numerical approximations. These are two completely new elements that we introduce to our PSD analysis. Note that the derivation of the distorted PSD, aims to pass all the windowing effects to the underlying fitted PSD model which has nothing to do with the smoothing window functions e.g.\ Parzen, Welch, Hann, Hamming etc., used in signal processing. The latter windows are used to reduce the power of side-lobs, away from the frequency of interest, and they are applied directly on the actual periodogram estimates i.e.\ they do not involve any knowledge of the actual underlying PSD model, in which most of the times is already known, in contrast to astrophysics which is the major unknown.\par
Thus, with our new approach all the spectral artefacts are correctly taken into account within the fitted model, and the probability density function for each Fourier bin, takes fully into account, through the convolution, the statistical dependencies from all the adjacent frequency bins. Thus, the estimation of confidence intervals for the best-fitting parameters can be carried out using the classical method of \citet{cash79} as described in the context of PSD, using the inverse of the generalised regularised incomplete gamma, in appendix~A2 in \citep{emmanoulopoulos13b}. Note, that for complicated functional form of the input PSD model the integrations, appearing in equation~\ref{eq:final_aver_peri}, may only be estimated numerically as the integrals may not have analytical solutions. The maximum likelihood approach is performed numerically (more details are given in EMM16) using classical nonlinear-global optimization procedures e.g.\ simulated annealing \citep{kirkpatrick83}.\par
As it is shown in EMM15, in the case of infinitely long and dense regular sampling, the method automatically (i.e.\ without the need to `switch something off'), converges to the classical maximum likelihood method as described in \citep[e.g.][]{anderson90, vaughan05, emmanoulopoulos13b}, since $\mathscr{P}_{\rm d}(f;\bmath{\gamma})$ goes to $\mathscr{P}(f;\bmath{\gamma})$ and thus, the d.o.f. at each Fourier frequency are equal to 2 (equation~\ref{eqe:dofDistorted}), yielding the classical $\chi^2$ distributions with 2 d.o.f, i.e.\ gamma distribution with shape parameter equal to one. In this limit case, the effects of `red-noise leak' and `aliasing' are completely absent.

\subsection{PSD analysis of \n7}
\label{ssect:psd_ngc7313}
In this section we perform a detailed PSD analysis of the \n7 light curves in different X-ray energy bands. Since the \textit{XMM-Newton} observations are practically accumulated in a continuous fashion, we are actually dealing with data averaged over time intervals equal to the sampling period, $t_{\rm bin}$, rather than simply sampled data, and thus the `aliasing effect' is not an issue \citep{vanderKlis88}. Moreover, our observations do not suffer from gaps, thus the window function is a continuous boxcar window function of finite length, equal to the duration of each of the four observations (Section~\ref{ssect:datRed}). Therefore, in the case of \n7 we are dealing only with the effect of `red-noise leak', due to the finite length of the window function. However, this effect is very marginal since, as we are going to see below in this section, all the PSD parameters (i.e.\ slopes and bends) are completely mapped within each observing window, at least for the first three observations 0725200301, 0725200101 and 0311190101, respectively.\par
Physically, for AGN variability studies, in which the PSD is described by a power-law with index is steeper than -1, the PSD should always become flat above a given time-scale, depicting the fact that the variability amplitude of the source can not be increasing for ever\footnote{When the power-law index is flatter than -1, the PSD can be integrated from zero frequency to any given frequency i.e. the variability amplitude could persist to arbitrarily lower frequencies without diverging.}. Note that for the commonly used power-law or broken power-law models, in order the variance of the underlying variability process not to be monotonically increasing as a function of time, the low frequency slope must be flatter than -1, and the high frequency slope must be steeper than -1 for the PSD to converge at low and high frequencies, respectively.\par 
Thus, the PSD model that we consider for \n7, is a stationary smoothly double bending power-law model, in which the stationarity property is introduced by fixing the values of low-frequency slope and low-frequency bend to 0 and $10^{-8}$ Hz, respectively. Thus, this assumption implies that the source does not exhibit larger amplitude variations on time-scales greater than 1157 days, since for larger time-scales (i.e.\ lower than $10^{-8}$ Hz) the PSD is characterised by a zero slope i.e.\ constant variance. This corresponds to an underlying variability process which is stationary for time-scales greater than $10^8$ s, or strictly speaking it is stationary up to second order \citep[e.g.][]{priestley81}, known also as weak/wide-sense stationarity. Note that there is robust observational evidence that the source does not exhibit longer time-scale variations based on long-term RXTE observations which consist of more than 870 days of observations covering sporadically the period between 1999 and 2001 (preliminary data analysis has been performed for another study). We have selected these values for both the low-frequency bend and slope since this is something which is expected to happen in AGN as shown by \citet{uttley02}, for the case of three AGN MCG-6-30-15, NGC\,5506 and NGC\,3516 exhibiting the same behaviour as the BHXRBs in the low state. The sensitivity of our best-fitting results to the actual value of these frozen two parameters is minuscule as discussed in Sect.~\ref{ssect:PSDestimComp}. 

Finally, we add to this PSD model a constant, $c$, that corresponds to the the Poisson noise level, and thus the overall model, $\mathscr{P}(f;\bmath{\gamma},c)$ can be written in the following form

\begin{figure}
\includegraphics[width=3.2in]{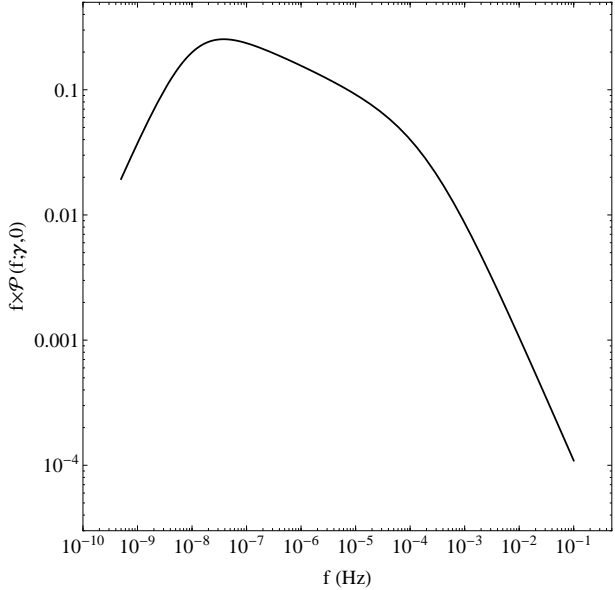}
\caption{An example of a double bending power-law PSD model with zero Poisson noise and model parameters 
$\bmath{\gamma}=\left\{0.01,0,1.2,2,10^{-8}\;{\rm Hz},2\times10^{-4}\;{\rm Hz}\right\}$}
\label{fig:modelPSD}
\end{figure}

\eqb
\mathscr{P}(f;\bmath{\gamma},c)=\frac{A f^{-\alpha_{\rm m}}}{\left(1+\left(\frac{f}{10^{-8}\;\rm{Hz}}\right)^{0-\alpha_{\rm m}}\right)\left(1+\left(\frac{f}{f_{\rm h}}\right)^{\alpha_{\rm h}-\alpha_{\rm m}}\right)}+c
\label{eqe:psd_bendModel}
\eqe
with $\bmath{\gamma}=\left\{A,\alpha_{\rm l}=0,\alpha_{\rm m},\alpha_{\rm h},f_{\rm l}=10^{-8}\;{\rm Hz},f_{\rm h}\right\}$, consisting of the normalization, low (fixed to 0), medium and high frequency slope and low (fixed to $10^{-8}$ Hz) and high-frequency bend, respectively. This PSD model is then inserted to equation~\ref{eq:final_aver_peri} yielding the distorted PSD model. Note that by fixing $\alpha_{\rm l}$ and $f_{\rm l}$ to the above-mentioned values, the PSD model normalization, $A$, corresponds to the power of the PSD model at $f_{\rm h}$ times $2{f_{\rm h}}^{\alpha_{\rm m}}\left(1+\left(f_{\rm h}/10^{-8}\right)^{-\alpha_{\rm m}}\right)$. Nevertheless, the normalization values that we quote throughout the manuscript are rescaled to the power of the PSD model at $f_{\rm h}$ times $2{f_{\rm h}}^{\alpha_{\rm m}}$, i.e.\ that of the single bending power-law model in which $f_{\rm l}=f_{\rm h}$, since this is the model that is currently used in the literature. In Fig.~\ref{fig:modelPSD} we show an example of a model PSD with zero Poisson noise having $\bmath{\gamma}=\left\{0.01,0,1.2,2,10^{-8}\;{\rm Hz},2\times10^{-4}\;{\rm Hz}\right\}$ and in order to visualize better the bend frequencies we multiply the PSD with the frequency. 

\subsection{PSD models in $0.5-10$ keV}
\label{ssect:psd0p5t10}
\subsubsection{Separate fits to individual observations: $\alpha_{\rm m}$ fixed to 1}
\label{ssect:separ_psd0p5t10fixAm}
\begin{figure*}
\includegraphics[width=3.4in]{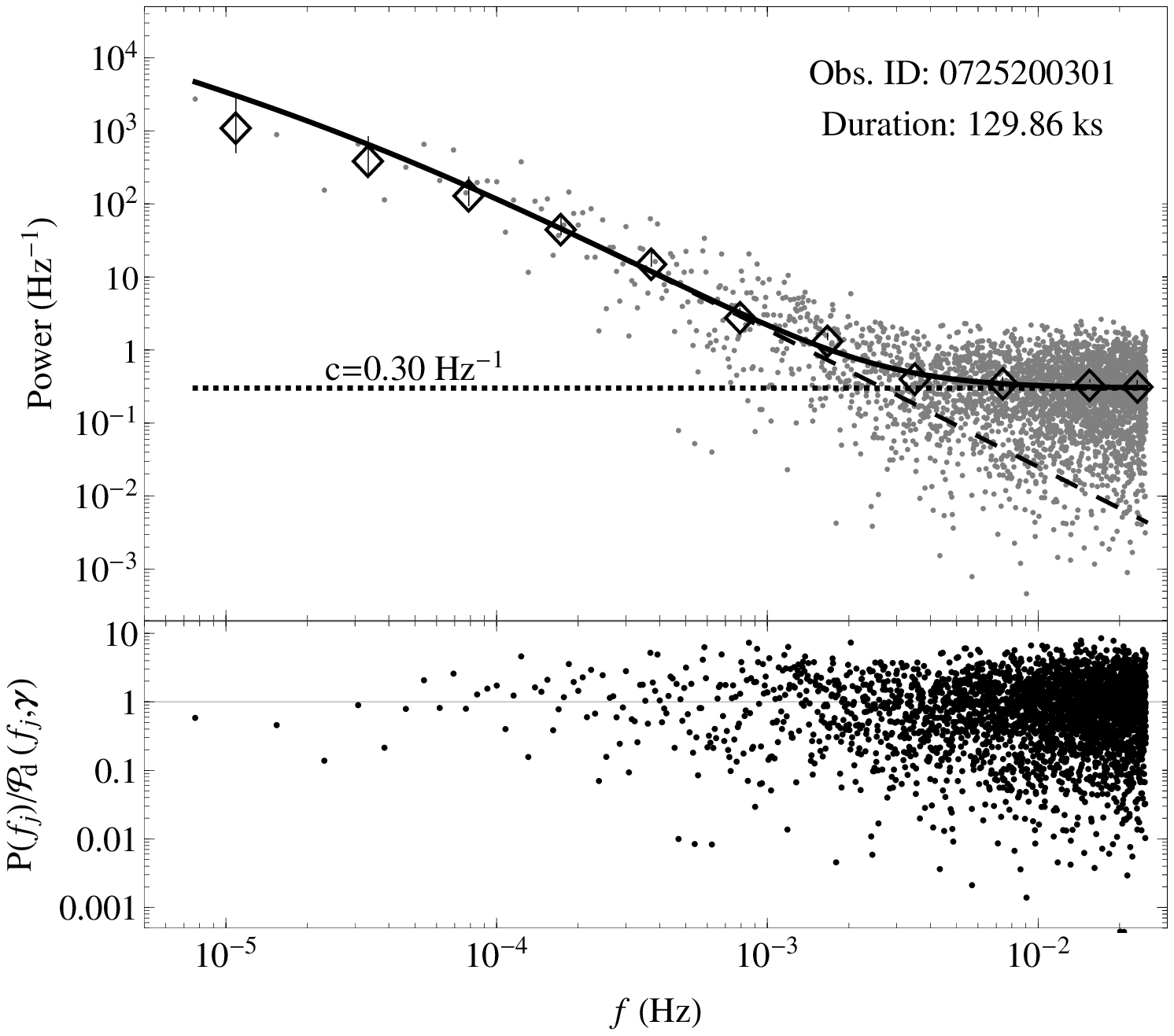}
\includegraphics[width=3.4in]{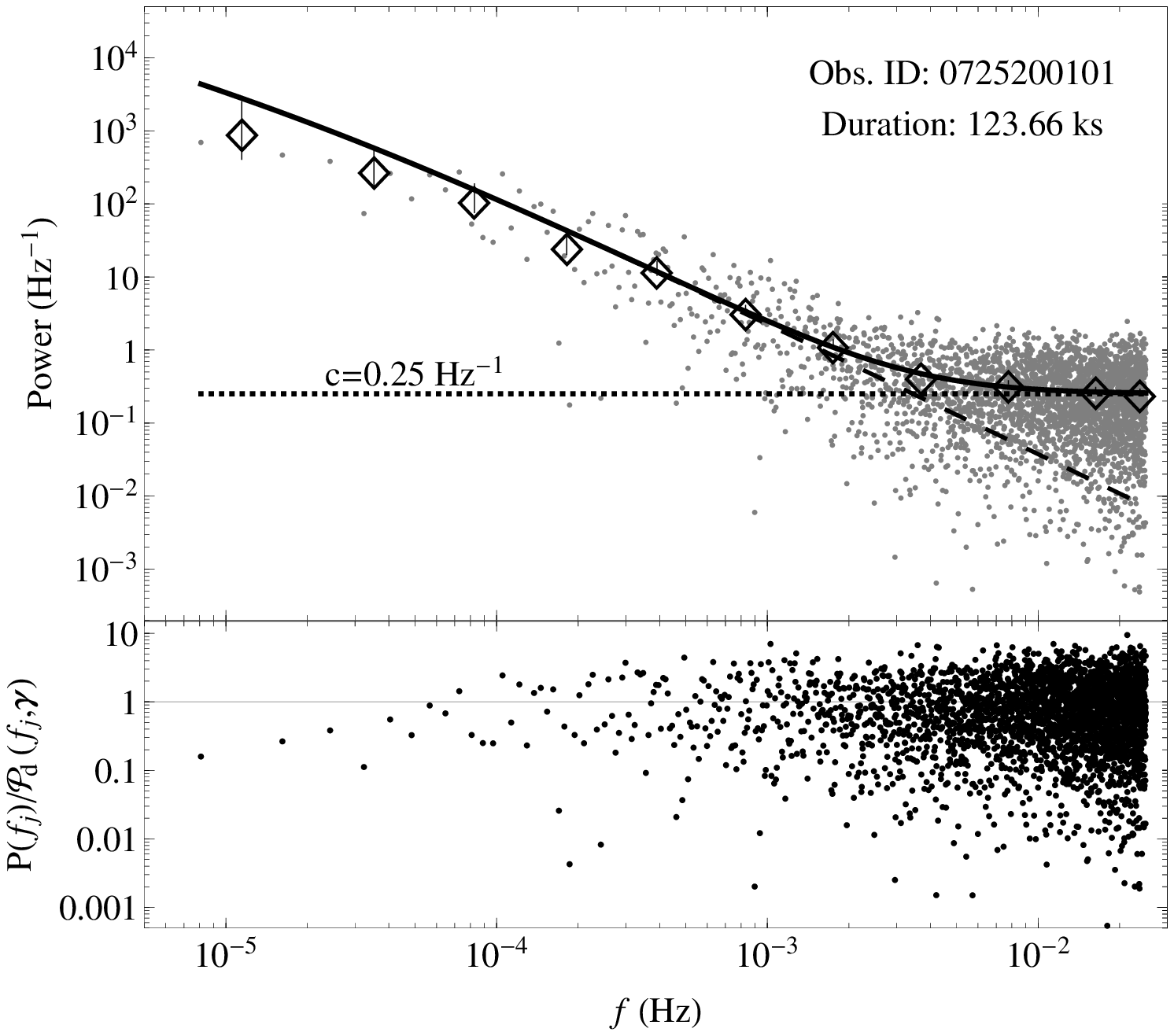}
\includegraphics[width=3.4in]{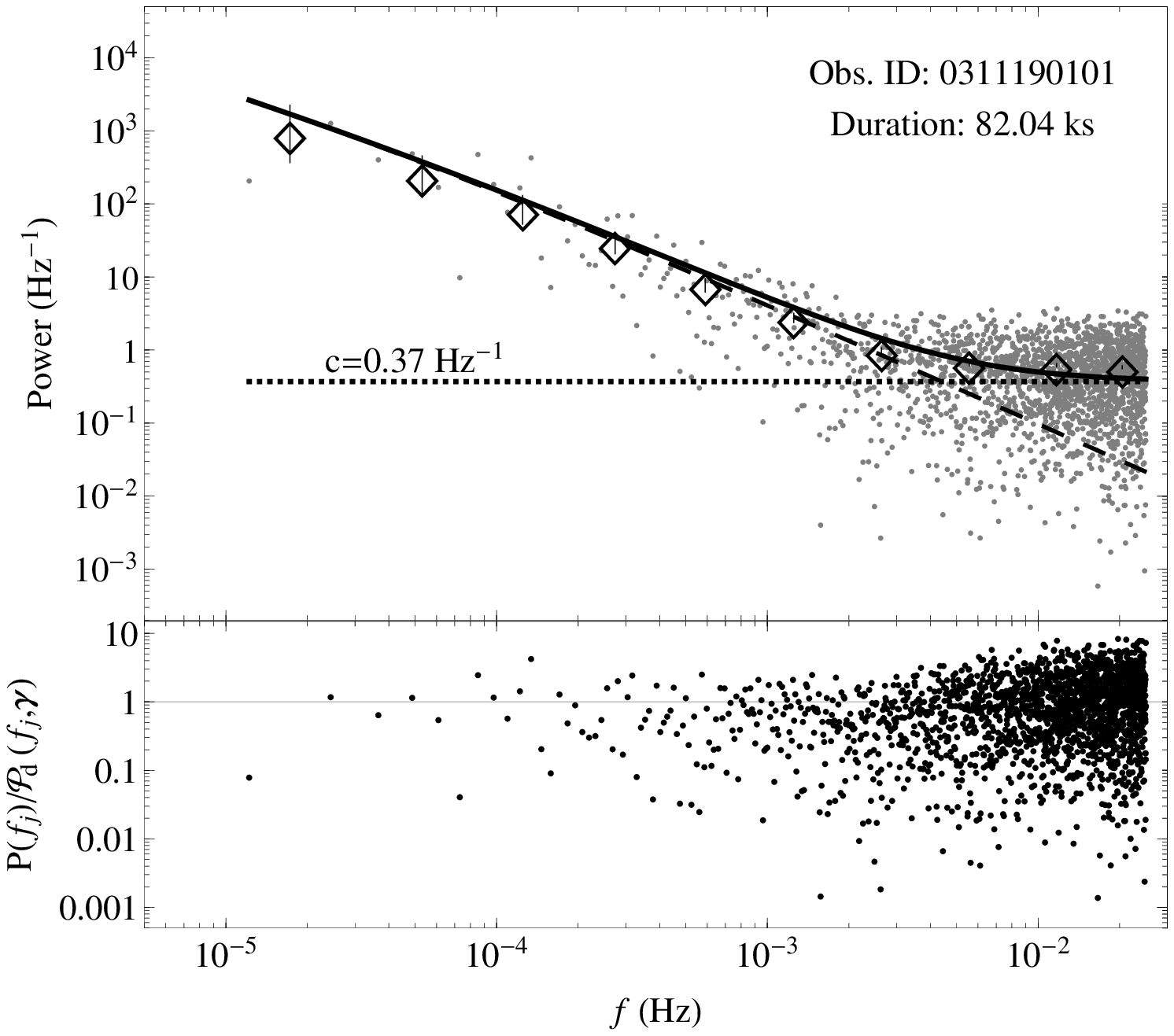}
\includegraphics[width=3.4in]{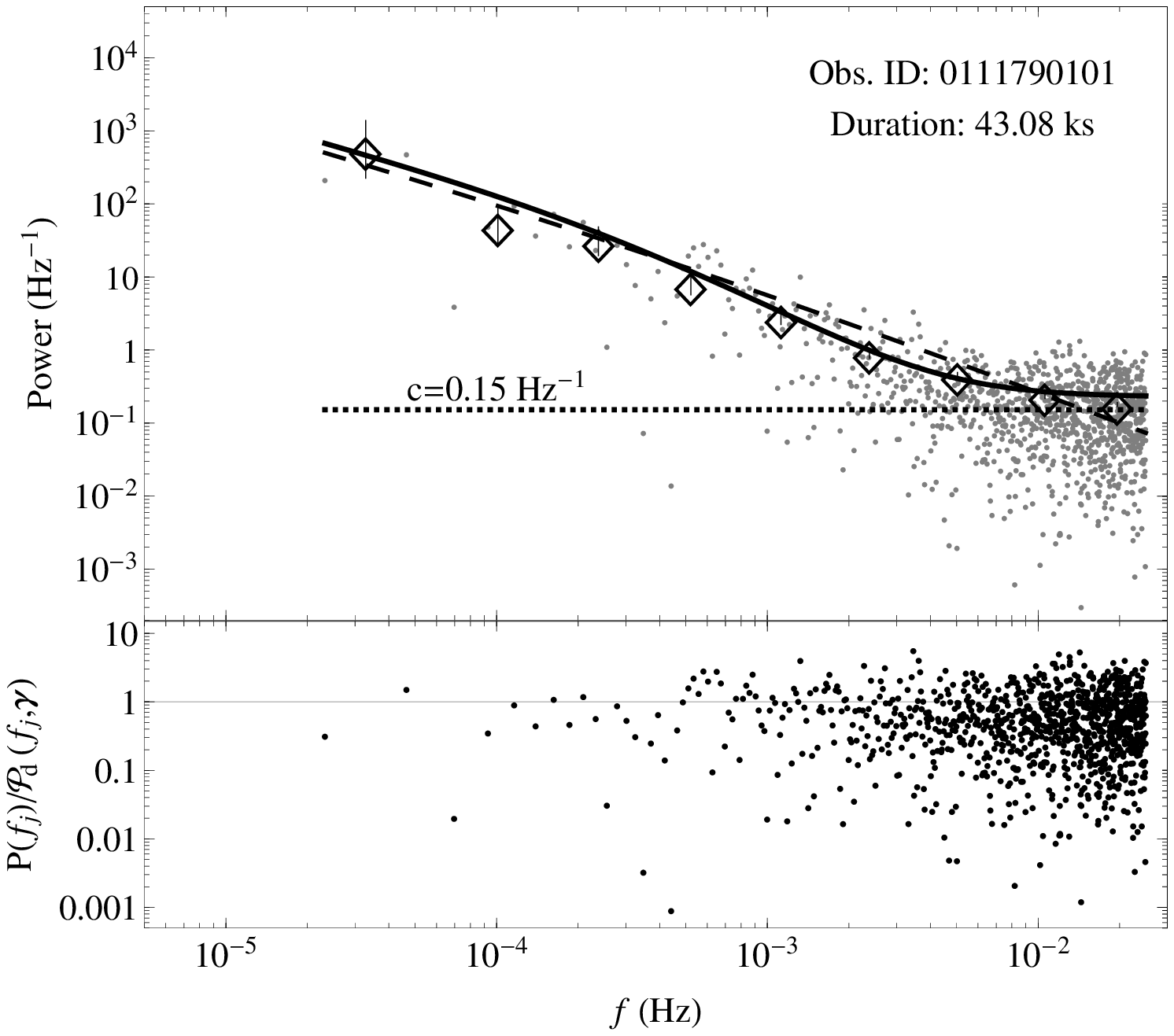}
\caption{The individual best-fitting PSD models, for the four observations considered here, in the $0.5-10$ keV energy range, with fixed value of $\alpha_{\rm m}$ to 1. Each panel shows for each observation the periodogram estimates, $P(f_j)$, with the grey points, the mean logarithmic periodogram estimates, with the open diamonds, and the best-fitting distorted PSD model, $\mathscr{P}_{\rm d}(f;\bmath{\gamma})$, with the solid line. The dashed line corresponds to the source's intrinsic best-fitting PSD model, $\mathscr{P}(f_j;\bmath{\gamma})$, and the horizontal dotted line to the Poisson noise level, $c$. The ratio plot, $P(f_j)/\mathscr{P}_{\rm d}(f_j;\bmath{\gamma})$, is attached to the bottom of each panel.}
\label{fig:psd_0.5_10}
\end{figure*}

Initially, we estimate the best-fitting double-bending PSD model, in the energy range of $0.5-10$ keV, for each observation separately. During the fit, we use the combined EPIC light curves and we fix the value of the medium-frequency slope, $\alpha_{\rm m}$, to unity \citep{markowitz03,mchardy06} since the \textit{XMM-Newton} data are considered to be relatively insensitive to both the low- (fixed to zero, for all our PSD analysis) and medium-frequency PSD model indices. In Fig.~\ref{fig:psd_0.5_10} we show for each observation the periodogram estimates, $P(f_j)$, (grey points) together with the mean logarithmic periodogram estimates \citep{papadakis93} (open diamonds) binned by a factor 2.1 in the logarithmic frequency space. The corresponding best-fitting distorted PSD model, $\mathscr{P}_{\rm d}(f;\bmath{\gamma})$ is shown by the solid line (equation~\ref{eq:final_aver_peri}) and carries all the spectral distortion effects i.e.\ in this case the red-noise leak. The underlying source best-fitting PSD model of \n7, $\mathscr{P}(f_j;\bmath{\gamma})$, is depicted by the dashed line, and the constant Poisson noise level, $c$ by the horizontal dotted line. At the bottom of each panel we show the ratio plot between the periodogram estimates at a given Fourier frequency, $P(f_j)$ over the best-fitting distorted PSD model, $\mathscr{P}_{\rm d}(f_j;\bmath{\gamma})$. The best-fitting PSD model parameter values are given in Table~\ref{tab:psd_0.5_10_am}.\par

\begin{table*}
\caption{Best-fitting source PSD model parameters for a smoothly double bending power-law model (equation~\ref{eqe:psd_bendModel}) to the individual observations (first column) in the $0.5-10$ keV energy range. Columns 2--5 show the best-fitting results having fixed $\alpha_{\rm m}$ to 1: the second column, (2), gives the high-frequency PSD bend, the third column, (3), the high-frequency PSD index, the forth column, (4), the PSD normalization, and the fifth column, (5), the constant Poisson noise level. Columns 6--10 show the best-fitting results having $\alpha_{\rm m}$ as a free fitting parameter: the sixth column, (6), gives the high-frequency PSD bend, the seventh column, (7), the medium-frequency PSD index, the eighth column (8), the high-frequency PSD index, the ninth column, (9), the PSD normalization, and the tenth column, (10), the constant Poisson noise level. Note that during the fits, $\alpha_{\rm l}$ and $f_{\rm l}$ are fixed to the values 0, and $10^{-8}$ Hz, respectively.}
\label{tab:psd_0.5_10_am}
\begin{tabular}{@{}cccccccccccc}
\hline
& \multicolumn{4}{c}{$\alpha_{\rm m}=1$} & & &\multicolumn{5}{c}{$\alpha_{\rm m}$ free fitting parameter}\\
(1) & (2) & (3) & (4) & (5) &  &  & (6) & (7) & (8) & (9) & (10)  \\
Obs. ID &  $f_{\rm h}\times10^{-5}$  & $\alpha_{\rm h}$ & $A\times10^{-2}$ & $c$ &  & & $f_{\rm h}\times10^{-5}$  & $\alpha_{\rm m}$ & $\alpha_{\rm h}$ & $A$ & $c$ \\
 & (Hz) &  & (Hz$^{-1}$) &  (Hz$^{-1}$) &  & & (Hz)  & & (Hz$^{-1}$) &  (Hz$^{-1}$) \\
\hline\hline
0725200301	&	$2.86^{+1.88}_{-1.05}$		&	$1.89^{+0.21}_{-0.14}$		&  $4.71^{+2.01}_{-1.98}$	& $0.30^{+0.25}_{-0.11}$ &  & &	$5.45^{+3.21}_{-2.27}$	&	$0.53^{+0.47}_{-0.32}$	& $2.15^{+0.53}_{-0.34}$	&	$61.26^{+18.56}_{-19.12}$ &	$0.35^{+0.26}_{-0.13}$\\  
0725200101	&	$2.01^{+1.81}_{-1.12}$		&	$1.79^{+0.24}_{-0.11}$		&  $5.23^{+1.91}_{-1.96}$	& $0.25^{+0.29}_{-0.06}$ & & &		$6.39^{+3.95}_{-2.45}$	&	$0.49^{+0.42}_{-0.34}$	& $1.98^{+0.46}_{-0.32}$	&	$72.21^{+21.59}_{-16.15}$ &	$0.29^{+0.30}_{-0.08}$\\  
0311190101	&	$2.18^{+1.92}_{-1.21}$		&  $1.65^{+0.28}_{-0.16}$		&  $5.60^{+2.08}_{-2.12}$	& $0.37^{+0.27}_{-0.09}$ & & & 	$7.12^{+4.09}_{-3.28}$	& $0.54^{+0.51}_{-0.35}$	& $2.08^{+0.48}_{-0.42}$	&  $52.98^{+18.79}_{-19.34}$	& $0.59^{+0.24}_{-0.11}$\\ 
0111790101	&	$29.07^{+3.23}_{-27.18}$	&  $2.02^{+0.96}_{-1.11}$		&  $1.53^{+5.12}_{-4.36}$	& $0.15^{+0.19}_{-0.05}$ & & & 	$13.24^{+4.56}_{-11.03}$&	$0.79^{+0.81}_{-\text{---}}$	& $2.17^{+1.06}_{-1.01}$	& 	$91.17^{+32.32}_{-27.22}$ & $0.17^{+0.18}_{-0.06}$\\ 
\hline
\end{tabular}
\end{table*}

The first two observations 0725200301 and 0725200101 are long enough in order the frequency bandpass to cover the high-frequency PSD bend, $f_{\rm h}$, at around $2\times10^{-5}$ Hz. Below this bend frequency there are still around 6 periodogram points resolving the actual form of the underlying PSD i.e.\ the value of $\alpha_{\rm m}$ which we have fixed to unity for this fit. This setting, $\alpha_{\rm m}=1$, causes the third observation, 0311190101, to yield a break at a similar frequency, despite the fact that the duration of this observation marginally matches the bend time-scale. The fourth observation, 0111790101, can not put any firm constraints to the break frequency, due to its limited length. However, all the observations yield a best-fitting high-frequency PSD slope, $\alpha_{\rm h}$, of around 1.85, with the weakest constrains coming, as it is expected, from the fourth observation.\par
The first two observations, due to their long duration, are not affected significantly from the red-noise leak effect, and thus for each case (Fig.~\ref{fig:psd_0.5_10}, top two panels) the best-fitting distorted PSD model (solid line) is effectively the sum of the underlying best-fitting PSD (dashed line) and the Poisson noise level (dotted line). However, the third observation is marginally affected from this spectral leakage, and as such the best-fitting distorted PSD model (Fig.~\ref{fig:psd_0.5_10}, bottom left-hand panel) carries in addition to the sum of the underlying best-fitting PSD and the Poisson noise level, the spectral leakage. Thus, the best-fitting high-frequency distorted PSD slope is flatter than that of the underlying PSD model and the flattening towards the high-frequency end of the spectrum is higher from the one expected purely by the Poisson noise level. Notice that in the previous two observations the solid and dashed lines are separated at around $10^{-3}$ Hz, in contrast to this observation in which the separation occurs at around few times $10^{-4}$ Hz.\par
Finally, the periodogram estimates of the forth observations are affected slightly more by the red-noise leak effect. Due to this additional variability power, that leaks from lower to higher frequencies, the flattening occurring at the high-frequency end of the spectrum (i.e.\ greater than $8\times10^{-3}$ Hz) sets at a relatively high power-level of around $0.25$ Hz$^{-1}$. Thus, the Poisson noise level for this data set appears to be the highest from all the other observations. However, by taking correctly the red-noise leak effect into account the actual best-fitting Poisson noise level is set to the lowest value, from the previous three observations, of $0.15$ Hz$^{-1}$, which is expected as this observation has the highest count-rate in the $0.5-10$ keV band (Fig.~\ref{fig:lcs}).\par
As we said before, for this fit $\alpha_{\rm m}$ is fixed to the widely used literature value of unity. Nevertheless, as we can see from the ratio plots of the first three longest observations, 0725200301, 0725200101 and 0311190101, below $10^{-4}$ Hz the medium-frequency slope is much flatter than 1. Note, that by fixing its value to unity affects also the derived value of the best-fitting values of $f_{\rm h}$ and $\alpha_{\rm h}$. Since, we do have significant information about the shape of the PSD at these low frequencies, from these three observations, in the next section we perform a joint fit to derive/constraint accurately and with the maximum possible precision the values of $\alpha_{\rm m}$, $\alpha_{\rm h}$ and $f_{\rm h}$.

\subsubsection{Separate fits to individual observations: $\alpha_{\rm m}$ free parameter}
\label{ssect:separ_psd0p5t10freeAm}
\begin{figure}
\includegraphics[trim = 0mm 7cm 0mm 6.5cm, clip, width=3.3in]{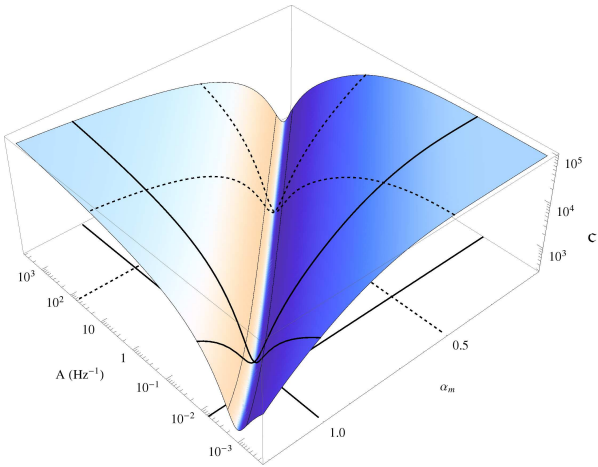}
\caption{The log-likelihood surface, $\mathcal{C}$, as a function of $\alpha_{\rm m}$ and $A$. The thin solid line, across the bottom of the surface, depicts the ridge of the minimum values of $\mathcal{C}$ for a given pair of $\alpha_{\rm m}$ and $A$. The solid thick contour lines correspond to the values of $\alpha_{\rm m}=1$ and $A=4.71\times10^{-2}$ Hz$^{-1}$, and the dashed contour lines correspond to the values of $\alpha_{\rm m}=0.5$ and $A=61.26$ Hz$^{-1}$.}
\label{fig:normVsMedSlope}
\end{figure}

In this section we fit to the periodograms of the individual observations the double-bending PSD model having $\alpha_{\rm m}$ as free model parameter. The best-fitting results are given in Table~\ref{tab:psd_0.5_10_am}. For all four observations the medium frequency slope, $\alpha_{\rm m}$, appears to be of the order of 0.5 and there is no evidence of variations in its value within the quoted uncertainties. Moreover, in comparison to the previous case, in which $\alpha_{\rm m}$ is fixed to unity, the best-fitting normalizations, $A$, are now almost 4 order of magnitudes larger. This is a very interesting point caused by the shape of the log-likelihood function, $\mathcal{C}$, in the $\alpha_{\rm m}$ versus $A$ parameter space. In Fig.~\ref{fig:normVsMedSlope} we show the form of this log-likelihood function for the case of the periodogram coming from obs ID: 0725200301 in the $0.5-10$ keV energy band. By setting the $\alpha_{\rm m}$ to be equal to 1 (solid contour line) the minimisation yields an $A$ of $4.71\times10^{-2}$ Hz$^{-1}$, whereas by setting the value to 0.5 (dashed contour line) we get a minimum at a value of $A$ of around 61.26 Hz$^{-1}$. As we are going to see in the next section (Sect.~\ref{ssect:psd0p5t10freeAm}), in which $\alpha_{\rm m}$ is free to vary, the latter set of values is the one which it is automatically chosen as the global minimum. The reason for underestimating the best-fitting normalisation parameter is to preserve the total variability power, since in this way it compensates for the `wrong' medium frequency power-law slope, which is fixed in the fitted PSD model to the value of 1. Thus, it is very crucial to leave both parameters free during the fit in order not to suppress unintentionally the normalization of the best-fitting PSD model.

\subsubsection{Joint fit to the ensemble of observations: $\alpha_{\rm m}$ free fitting parameter}
\label{ssect:psd0p5t10freeAm}
\begin{figure*}
\includegraphics[width=2.3in]{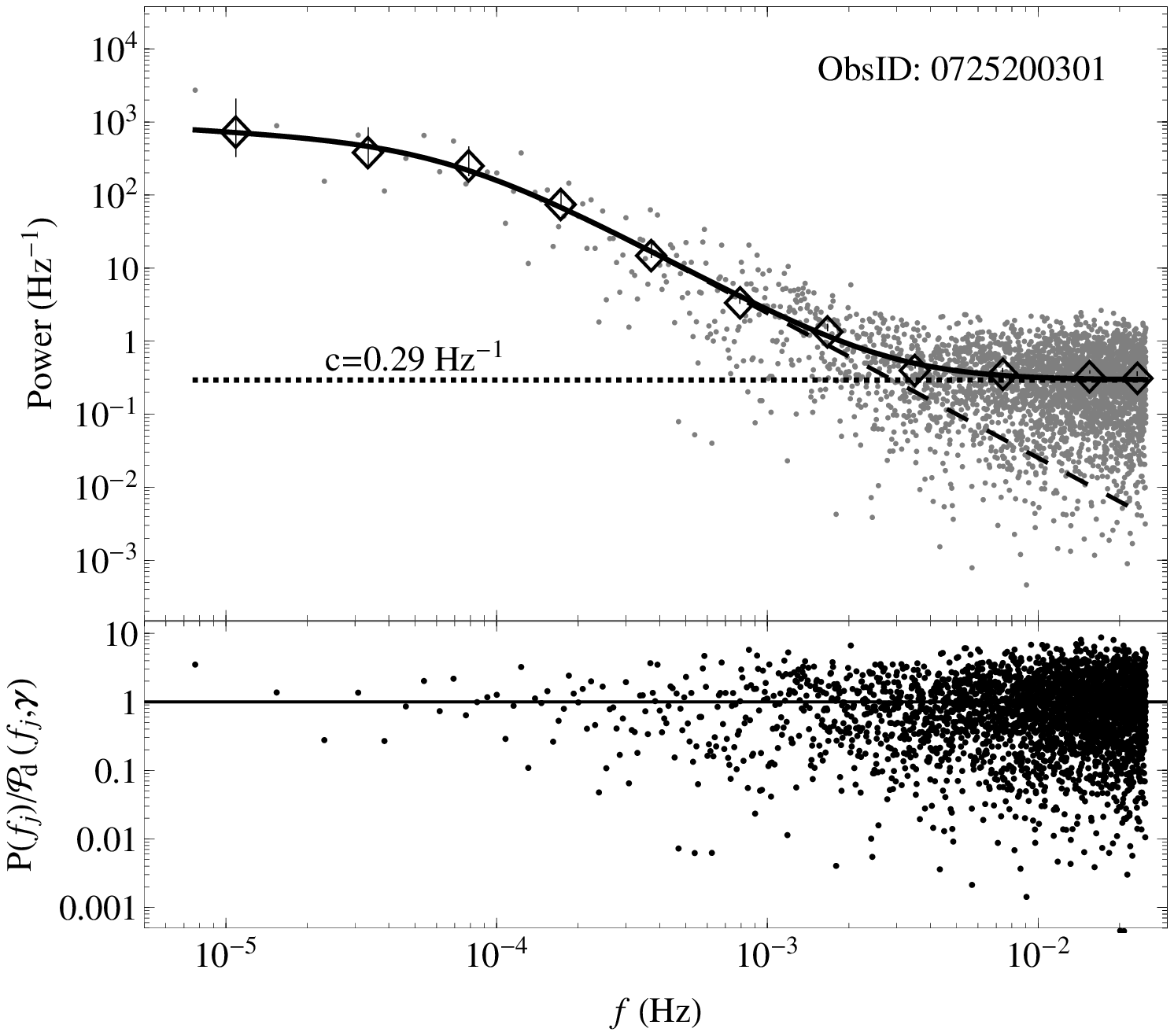}
\includegraphics[width=2.3in]{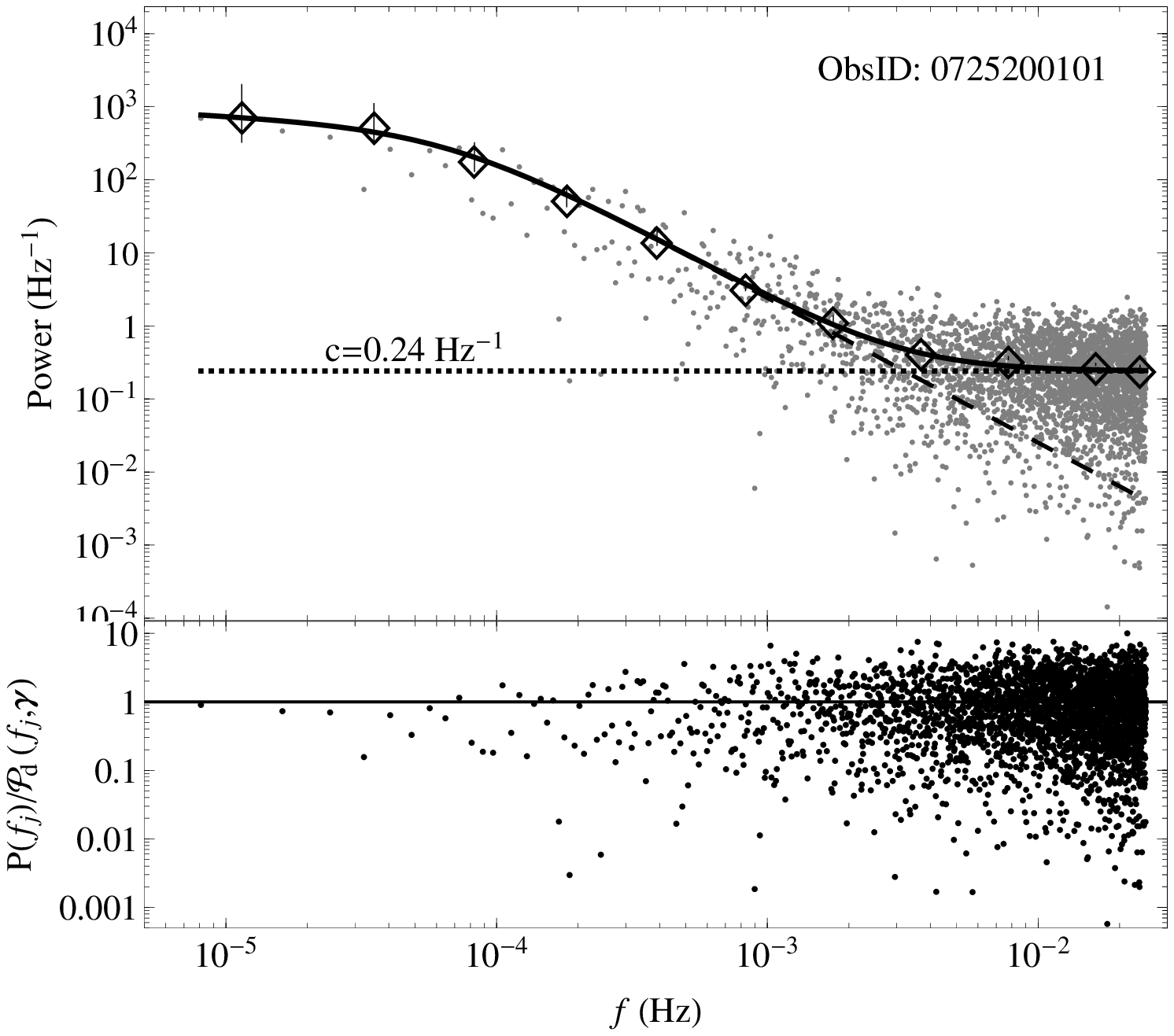}
\includegraphics[width=2.3in]{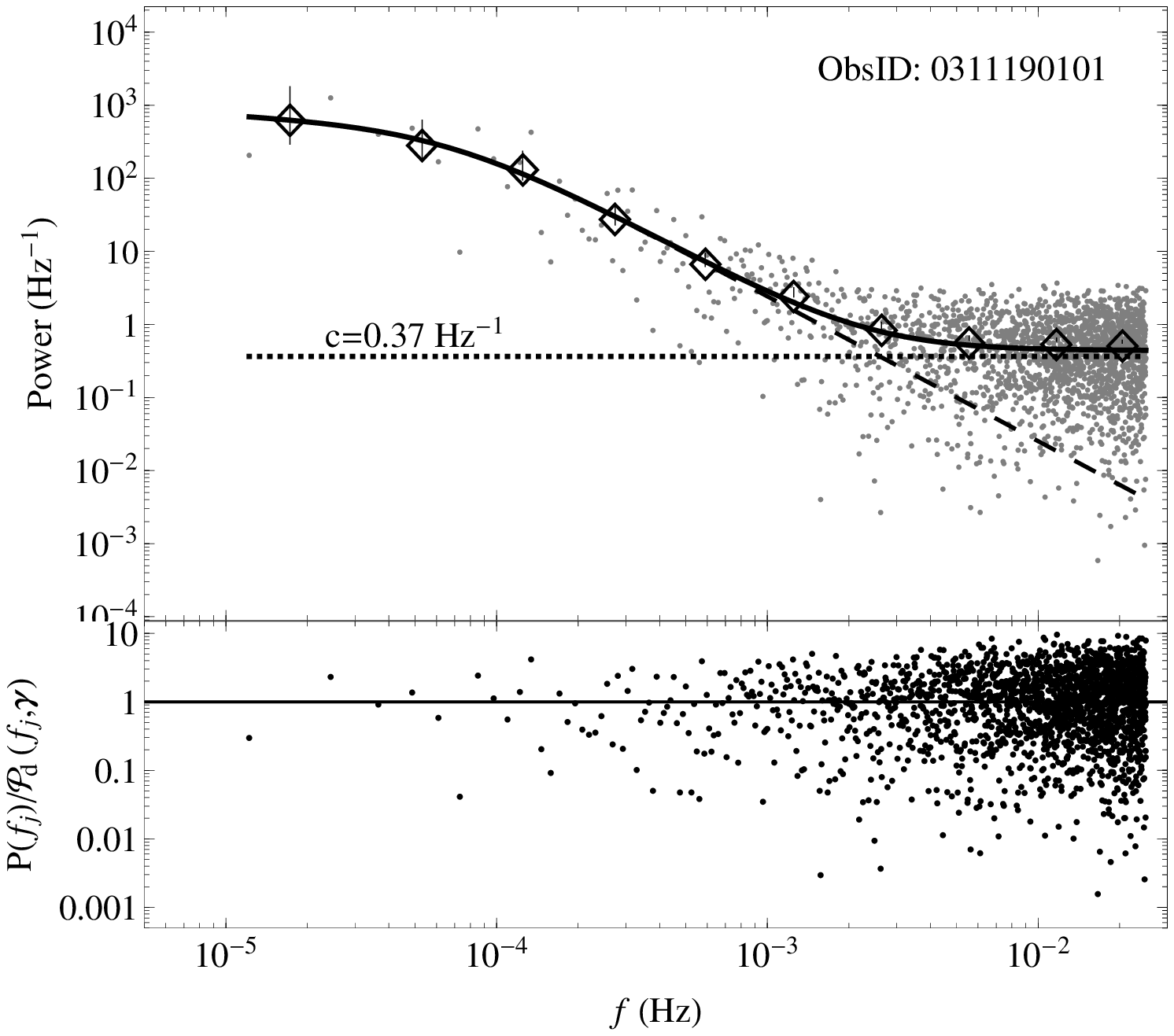}
\caption{The joint best-fitting PSD model in the $0.5-10$ keV energy range. Each panel shows for each observation the periodogram estimates, $P(f_j)$, with the grey points, the mean logarithmic periodogram estimates, with the open diamonds, and the best-fitting distorted PSD model, $\mathscr{P}_{\rm d}(f;\bmath{\gamma})$, with the solid line. The dashed line corresponds to the source's intrinsic best-fitting PSD model, $\mathscr{P}(f_j;\bmath{\gamma})$, and the horizontal dotted line to the Poisson noise level, $c$. The ratio plot, $P(f_j)/\mathscr{P}_{\rm d}(f_j;\bmath{\gamma})$, is attached to the bottom of each panel.}
\label{fig:psdComp_0.5_10Joint}
\end{figure*}

\begin{table*}
\caption{Best-fitting source PSD model parameters for the joint fit of a smoothly double bending power-law model (equation~\ref{eqe:psd_bendModel}) to the three longest observations: 0725200301, 0725200101 and 0311190101. The first column, (1), gives the energy range, the second column, (2), the high-frequency PSD bend, the third column, (3), the medium-frequency PSD index, the forth column, (4), the high-frequency PSD index, the fifth column, (5), the PSD normalization, and finally the sixth, seventh and eighth column, (6,7 and 8), the constant Poisson noise levels for the three observations, respectively. Note that during the fit, $\alpha_{\rm l}$ and $f_{\rm l}$ are fixed to the values 0 and $10^{-8}$ Hz, respectively.}
\label{tab:joint_psd_energy_bands}
\begin{tabular}{@{}cccccccc}
\hline
(1) & (2) & (3) & (4) & (5) & (6) & (7) & (8)\\
Energy range &  $f_{\rm h}\times10^{-5}$  & $\alpha_{\rm m}$ & $\alpha_{\rm h}$ & $A$ & $c_1$ & $c_2$ & $c_3$ \\
(keV) &(Hz) & &  &  (Hz$^{-1}$) & (Hz$^{-1}$) & (Hz$^{-1}$) & (Hz$^{-1}$)\\
\hline\hline
$0.5-10$	& $6.71^{+1.31}_{-0.97}$	& $0.51^{+0.30}_{-0.19}$	& $1.99^{+0.17}_{-0.06}$	& $79.64^{+5.96}_{-3.73}$	& $0.29^{+0.13}_{-0.06}$	& $0.24^{+0.18}_{-0.05}$		& $0.37^{+0.19}_{-0.05}$	\\  
$0.5-2$		& $6.05^{+1.89}_{-1.06}$	& $0.32^{+0.48}_{-0.24}$	& $2.21^{+0.22}_{-0.10}$	& $41.23^{+7.43}_{-5.06}$	& $0.67^{+0.99}_{-0.12}$	& $0.55^{+1.06}_{-0.33}$		& $0.86^{+1.11}_{-0.12}$	\\  
$2-4$		& $4.93^{+1.92}_{-1.39}$	& $0.75^{+0.67}_{-0.39}$	& $1.86^{+0.25}_{-0.12}$	& $243.01^{+10.04}_{-9.39}$	& $0.95^{+0.96}_{-0.26}$	& $0.77^{+1.01}_{-0.31}$		& $1.21^{+1.12}_{-0.45}$	\\ 
$4-10$		& $4.26^{+2.08}_{-\text{---}}$	& $0.83^{+0.74}_{-0.42}$	& $1.78^{+0.34}_{-0.22}$	& $689.64^{+15.98}_{-12.34}$	& $1.47^{+1.10}_{-0.12}$	& $1.22^{+1.07}_{-0.11}$		& $1.80^{+1.23}_{-0.15}$	\\
\hline
\end{tabular}
\end{table*}
In this section, we estimate the best-fitting double-bending PSD model, again in the energy range of $0.5-10$ keV, but this time we perform a joint fit using the first three longest observations, 0725200301, 0725200101 and 0311190101. This allows us to leave the medium-frequency slope, $\alpha_{\rm m}$, as a free fitting parameter since it sampled adequately by the corresponding periodogram estimates. All the PSD model parameters i.e. $\bmath{\gamma}=\left\{A,\alpha_{\rm l}=0,\alpha_{\rm m},\alpha_{\rm h},f_{\rm l}=10^{-8}\;{\rm Hz},f_{\rm h}\right\}$ are tied together and the Poisson noise levels are allowed to vary freely for each observation separately\footnote{Note that throughout this work (as we write in Section~\ref{ssect:psd_ngc7313}) the low-frequency slope, $\alpha_{\rm l}$, and the low-frequency bend frequency, $f_{\rm l}$, continue to be fixed to 0 and $10^{-8}$ Hz, respectively.}. The joint best-fitting distorted and intrinsic PSD models are shown, separately for each observation, in Fig.~\ref{fig:psdComp_0.5_10Joint}. The best-fitting source model parameters are given in Table~\ref{tab:joint_psd_energy_bands} and they are: normalization, $A=79.64$ Hz$^{-1}$, high-frequency PSD bend, $f_{\rm h}=6.71\times10^{-5}$ Hz, medium and high PSD indices, $\alpha_{\rm m}=0.51$ and $\alpha_{\rm h}=1.99$, and the Poisson levels for the three observations, $c_1=0.29$, $c_2=0.24$ and $c_3=0.37$ Hz$^{-1}$, respectively. In comparison to the previous best-fitting results, for which $\alpha_{\rm m}$ was fixed to unity (as it is usually assumed in the literature for AGN variability studies), the actual value of the $\alpha_{\rm m}$ is half from it, and the break frequency is now shifted almost a factor of three towards the higher frequencies.

Adding the fourth observation (being the shortest), 0111790101, to the overall fitting process, makes the global optimisation process very cumbersome. Based on the last best-fitting results, neither the position of $f_{\rm h}$ nor the range of $\alpha_{\rm m}$ are sampled by the periodogram estimates of this observation. This causes a great deal of problems, not only during the estimation of the best-fitting parameters, but also during the estimation of uncertainties that almost in all cases yields upper/lower limits. Thus, from the following PSD analysis we exclude this observation from our computations. In Appendix~\ref{app1:confidence_limits} we show the form of the log-likelihood functions that we used to derive the the 68.3 per cent confidence limits and we also give the joint confidence limits for $f_{\rm h}$ and $\alpha_{\rm h}$, respectively.

\subsection{PSD models in $0.5-2$, $2-4$ and $4-10$ keV}
In this section, we study the shape of the underlying PSD model of \n7 at different energy bands: $0.5-2$, $2-4$ and $4-10$ keV. For this study we use only the first three observations, 0725200301, 0725200101 and 0311190101, which are the longest ones, and thus able to disclose possible differences in the position of the high-frequency slope $f_{\rm h}$. For each energy band we perform a joint fit, exactly as we did previously for the periodogram estimates in the $0.5-10$ keV energy band.\par
\begin{figure*}
\includegraphics[width=3.36in]{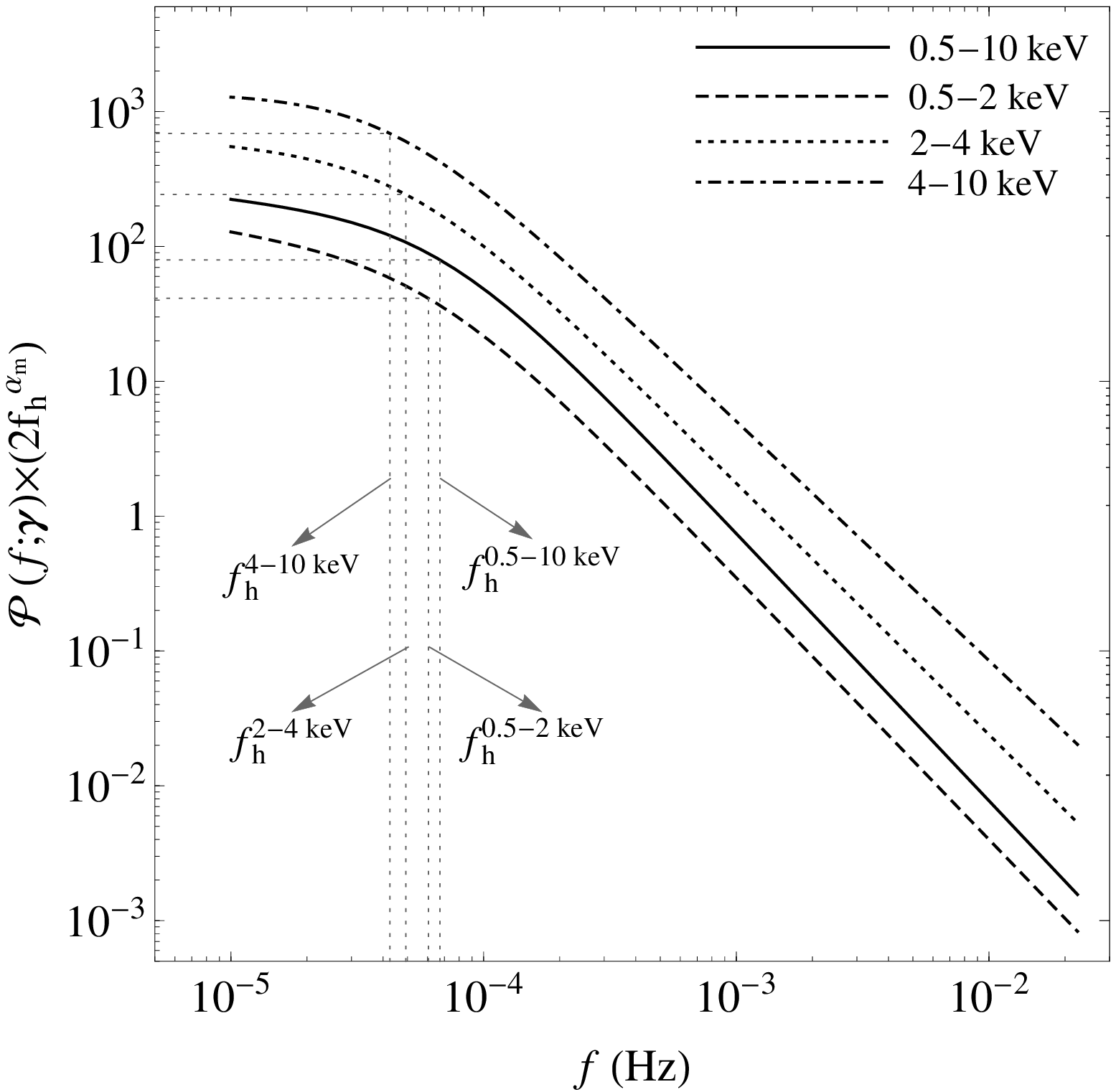}\hspace{1em}
\includegraphics[width=3.3in]{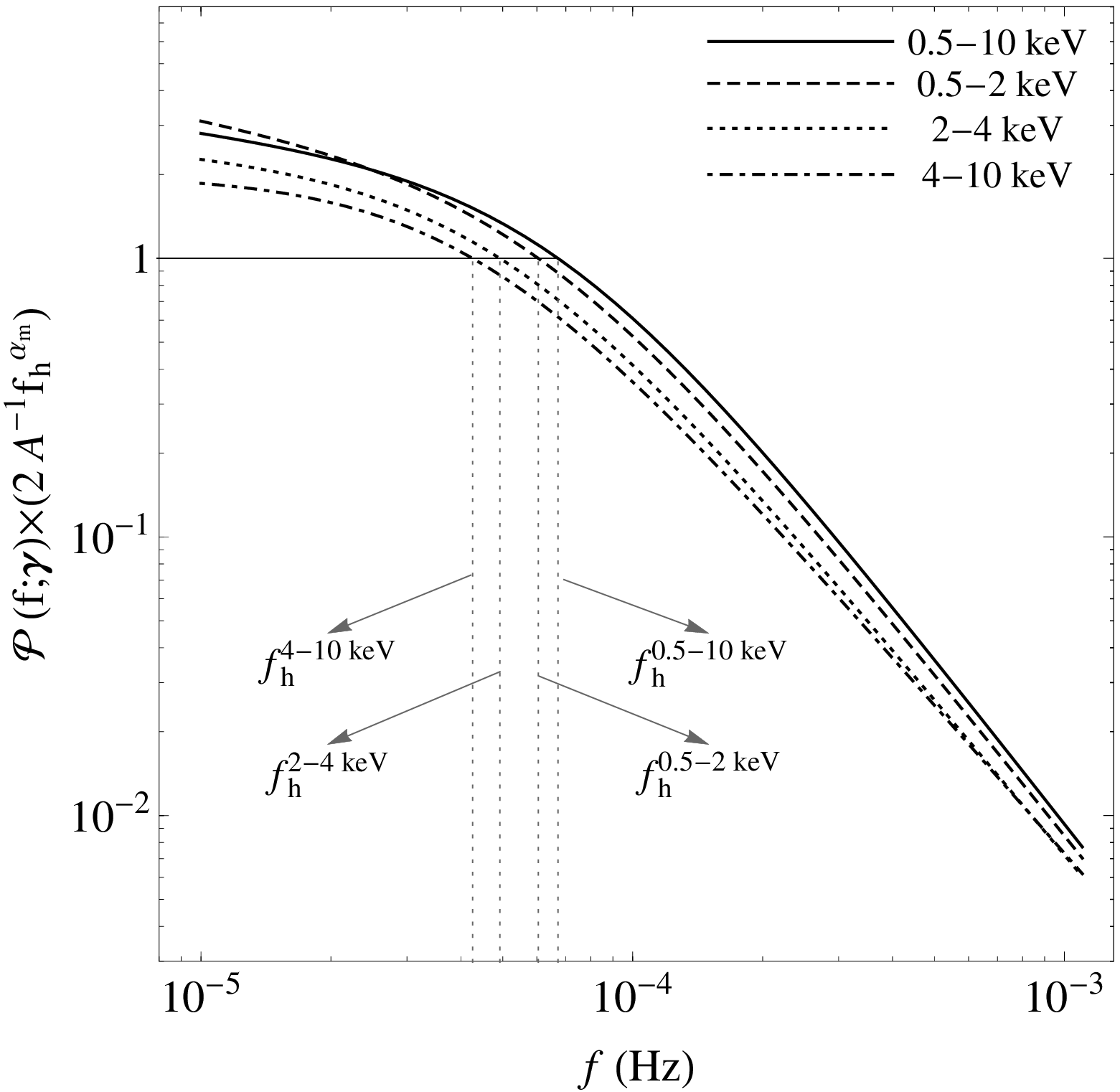}
\caption{Best-fitting source PSD models as a function of energy. In both panels, the grey dotted lines show the positions of $f_{\rm h}$ and $A$. Left-hand panel: The best-fitting PSD models are multiplied by $2{f_{\rm h}}^{\alpha_{\rm m}}$, in order the normalization to correspond to the variability power at $f_{\rm h}$. This panel shows that the higher the energy range the larger the normalization, $A$. Right-hand panel: The best-fitting PSD models have been multiplied by $2A^{-1}{f_{\rm h}}^{\alpha_{\rm m}}$, in order the PSD value at the break frequency to be equal to unity, $\mathscr{P}(f_{\rm h};\bmath{\gamma})=1$. This panel shows that the higher the energy range the flatter the high-frequency PSD slope, $\alpha_{\rm h}$. Note that if we ignore the parameter uncertainties for both $f_{\rm h}$ and $\alpha_{\rm m}$ parameters, both the plots support the hint that as the energy increases $f_{\rm h}$ moves to lower frequencies and $\alpha_{\rm m}$ becomes steeper.}
\label{fig:energDependPSDparam}
\end{figure*}

Based on the best-fitting source PSD model parameter values, given in Table~\ref{tab:joint_psd_energy_bands}, we can clearly see that as the energy increases the high-frequency PSD slopes, $\alpha_{\rm h}$, become flatter and the PSD model normalizations, $A$, become larger. In the left-hand panel of Fig.~\ref{fig:energDependPSDparam} we show the best-fitting source PSD models, for each energy range, multiplied by the quantity $2f_{\rm h}^{\alpha_{\rm m}}$ (see Section~\ref{ssect:psd_ngc7313}) in order the normalization to correspond to the PSD value at $f_{\rm h}$, i.e.\ $A=\mathscr{P}(f_{\rm h};\bmath{\gamma})$. Moreover, there are some hints that as the energy increases, the medium-frequency PSD slopes, $\alpha_{\rm m}$, become steeper and that the high-frequency PSD bends, $f_{\rm h}$, are shifted towards smaller frequencies. These hints are based only on their best-fitting values and within the derived 68 per cent confidence limits we can not statistically support these two tendencies. Both of these 
features can also be seen in the previous plot, but in order to visualize them properly we multiply the PSD best-fitting models by the factor $2A^{-1}f_{\rm h}^{\alpha_{\rm m}}$ and the PSD value at the break frequency becomes equal to unity, $\mathscr{P}(f_{\rm h};\bmath{\gamma})=1$ (Fig.~\ref{fig:energDependPSDparam}, right-hand panel).

\subsection{PSD estimation: Comparison with the classical method}
\label{ssect:PSDestimComp}
A very important point is that for the \textit{XMM-Newton} observations of \n7, that we are using in this paper, both the low-frequency part of the PSD, as well as the medium-frequency bend, are well sampled by our periodogram estimates, and thus the effect of red-noise leak is almost non-existent. If we repeat the PSD fitting procedure of Sect.~\ref{ssect:psd0p5t10freeAm} (i.e.\ ensemble of observations in the $0.5-10$ keV, having $\alpha_{\rm m}$ as a free fitting parameter) but this time we use the simple single-bending power-law model, $\mathscr{P}(f;\bmath{\gamma})$ equation~\ref{eqe:psd_bendModel}, by simply replacing $\alpha_{\rm l}$ with $\alpha_{\rm m}$ (i.e.\ $\bmath{\gamma}=\{A,f_{\rm h}, \alpha_{\rm m}, \alpha_{\rm h}\}$), i.e.\ vanishing the first multiplicative term in the parentheses in the denominator. Then, we fit this PSD model directly to the periodogram estimates, without performing the convolution with the window function (equation~\ref{eq:final_aver_peri}). For the fitting procedure we use the classical maximum likelihood approach described in appendix~A2 in \citet{emmanoulopoulos13b} based on the gamma-distribution. The best-fitting PSD model parameters that we get are $\bmath{\gamma}_{\rm bf}=\left\{71.56^{+8.91}_{-6.89}\right.$ Hz$^{-1}$, $\left(4.96^{+1.69}_{-1.33}\right)\times10^{-5}$ Hz, $0.58^{+0.28}_{-0.24}$,$\left.2.02^{+0.19}_{-0.11}\right\}$ which are completely consistent, as expected, with the best-fitting results coming from our new method (Table~\ref{tab:joint_psd_energy_bands}). As we can see from Fig.~\ref{fig:psd_0.5_10}, except from the case of the shortest observation (obs. ID: 0111790101, lower right-hand panel), for all the other observations the intrinsic source spectrum (dashed line) plus the Poisson noise level is in accordance with the distorted PSD model (solid line) dictating that by fitting a simple bending power-law model directly to the periodograms, will yield a correct estimate of the intrinsic PSD of the source.\par
Finally, as we said in Sect.~\ref{ssect:psd_ngc7313} during our PSD modelling, using the distorted PSD, we fix the values of the low-frequency slope and bend, $\alpha_{\rm l}$ and $f_{\rm l}$, to the values 0 and $10^{-8}$ Hz, respectively. By changing the $f_{\rm l}$ by plus/minus two orders of magnitude, the best-fitting results remain the same (within the quoted uncertainties). Similarly as before, by considering four different low frequency slopes between 0 and 1 (0.2, 0.4, 0.6 and 0.8) the results remain the same. Note, that the stationarity property, expressed through freezing $\alpha_{\rm l}$ and $f_{\rm l}$, has only been introduced for consistency reasons. Our \textit{XMM-Newton} light curves, lasting around 1.5 days, do not exhibit any sort of long-term trends, thus the effect of red-noise leakage is very minimal. In the forthcoming paper (EMM16) we show that for the limit case of uniform sampling, in the absence of red-noise leak and spectral aliasing, the distorted periodogram (equation~\ref{eq:final_aver_peri}) is identical to the underlying PSD model, $\mathscr{P}_{\rm d}(f;\bmath{\gamma})\equiv\mathscr{P}(f;\bmath{\gamma})$.

\section{PSD SCALING RELATIONS}
\label{sect:scale_relations}

\begin{figure*}
\includegraphics[width=3.3in]{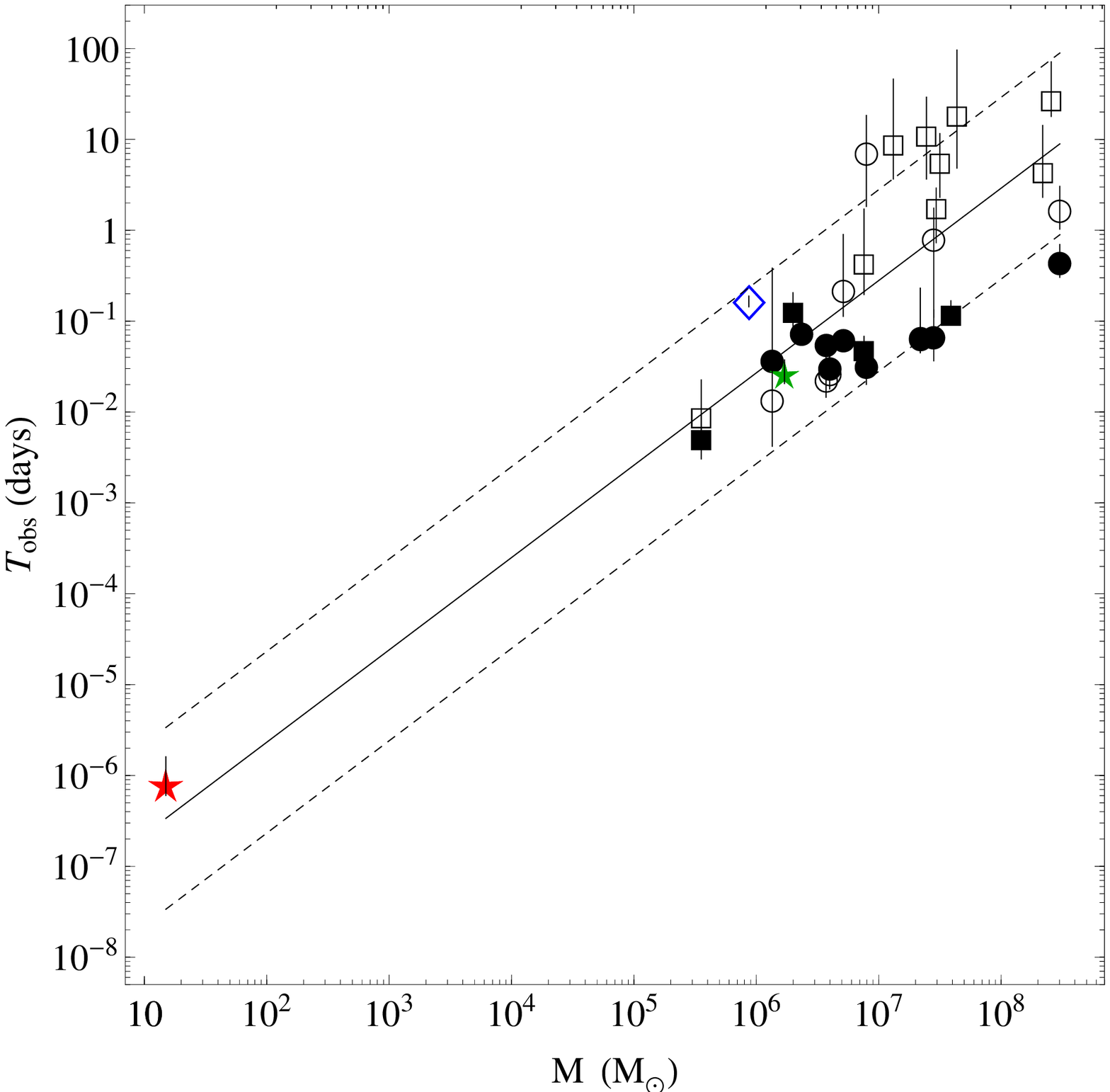}\hspace{2em}
\includegraphics[width=3.3in]{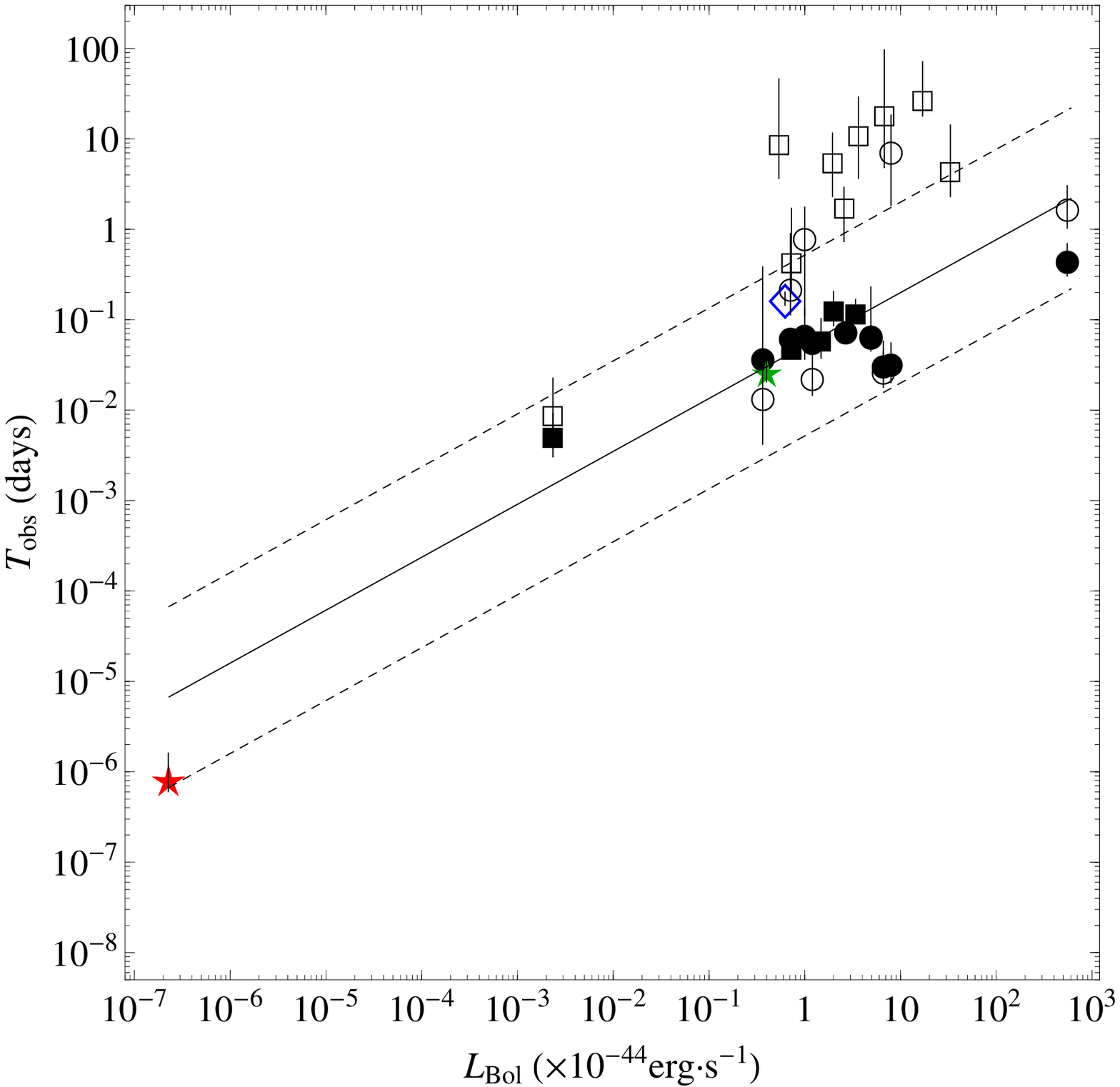}
\caption{The linear dependences between $T_{\rm obs}$ and $M$ and $L_{\rm bol}$. \n7 is depicted by the open blue diamond. The rest of the points correspond to the GV12 AGN sample: circles represent narrow-line Seyfert 1, squares represent type-1 Seyferts, the small green star corresponds to the single type-2 Seyfert and the red star depicts Cygnus X-1. The open symbols are data-points reported in the literature and the filled symbols are the sources analysed in GV12. [Left-hand panel] $T_{\rm obs}$ versus $M$: The solid line corresponds to the best-fitting linear model, described by equation~\ref{eq:MbhTb} with $A=1.02\pm0.21$ and $C=-1.57\pm0.27$. The dashed lines illustrate the one-order of magnitude ($\dex$) regions above and below the best-fitting model. [Right-hand panel] The solid line corresponds to the best-fitting linear model, described by equation~\ref{eq:TbLbol} with $A=0.59\pm0.22$ and $C=-0.70\pm0.22$. The dashed lines illustrate the one-order of magnitude ($\dex$) regions above and below the best-fitting model.}
\label{fig:MbhTbLbol}
\end{figure*}

In this section we consider how our results might affect the scaling relations reported by \citet{gonzalez12} (GV12, hearafter) in which PSD bend time-scales, derived just from \textit{XMM}-\textit{Newton} observations. We note that we are only adding one more point to the GV12 ensemble of sources and, although the bend frequency of \n7 is well measured in our work, the BH mass is estimated from stellar velocity dispersion, rather than optical reverberation (see Sect.~\ref{sect:intro}), and so is subject to a larger uncertainty. For consistency purposes, we repeat exactly the same fitting procedure\footnote{We use the very simple unweighed least square technique. The reasons are explained in footnote 8 in GV12.} presented in GV12, adding to their sample our estimates PSD bend time scale estimate of \n7. The errors in the best-fitting parameters are the standard errors of the linear regression analysis. 

Our best-fitting high-frequency PSD bend frequency in $0.5-10$ keV energy band is $1.94\times10^{-4}$ Hz (Section~\ref{ssect:psd0p5t10freeAm}), the BH mass and the bolometric luminosity of the sources is $0.87\times10^6$ \ms\ and $(6.21\pm3.47)\times10^{43}$ erg s$^{-1}$, respectively (Section~\ref{sect:intro}). Initially, we assume that the observed bend time-scale $T_{\rm obs}=f_{\rm h}^{-1}$ depends linearly only on and the BH mass, $M$, thus, in the logarithmic space, this dependence can be written as

\eqb
\log_{10}[T_{\rm obs}]=A\log_{10}[M]+C
\label{eq:MbhTb}
\eqe

The best-fitting model has $A=1.02\pm0.21$ and $C=-1.57\pm0.27$ with a sum of squared errors of 11.99 for 20 degrees of freedom. The best-fitting model is shown in the left panel of Fig.~\ref{fig:MbhTbLbol} and the open-blue diamond corresponds to \n7.\par
Then, we consider a model in which the bend time-scale, $T_{\rm obs}$, has an additional linear dependence on the bolometric luminosity, $L_{\rm bol}$,

\eqb
\log_{10}[T_{\rm obs}]=A\log_{10}[M]+B\log_{10}[L_{\rm bol}]+C
\label{eq:MbhTbLbol}
\eqe

The best-fitting model has $A=1.18\pm0.34$, $B=-0.17\pm0.28$ and $C=-1.69\pm0.34$ with a sum of squared errors of 11.76 for 19 degrees of freedom. 

Our results for both models agree fully with those reported by GV12 indicating, that there is not a statistical need to introduce a bolometric luminosity dependence to the high-frequency bend time-scale. The dependence of $T_{\rm obs}$ on $L_{\rm bol}$ is governed by the best-fitting parameter $B$, which from our analysis is consistent with zero, since it has a null hypothesis probability of 0.55.\par

Finally, we consider a linear model between the bend-frequency and the luminosity of the following form

\eqb
\log_{10}[T_{\rm obs}]=A\log_{10}[L_{\rm bol}]+C
\label{eq:TbLbol}
\eqe

and we fit it to the GV12 sample including, similarly as before, our PSD bend time scale for \n7. The best-fitting model is shown in the right-hand panel of Fig.~\ref{fig:MbhTbLbol} and the best-fitting model parameters are $A=0.59\pm0.22$ and $c=-0.70\pm0.22$ yielding a sum of squared errors 19.064 for 20 degrees of freedom.

\section{TIME-LAG SPECTRA}
\label{sect:tlspec}
\subsection{Estimation of the time-lag spectra}
\label{ssect:TL_data_estim}
In order to estimate the time-lag spectrum between two light curves we use the standard analysis method outlined in \citet{bendat86,nowak99}. In brief, consider for a given source a soft and a hard light curve, $s(t)$ and $h(t)$ respectively, obtained simultaneously and consisting of the same number of $N$ equidistant observations with a sampling period $t_{\rm bin}$. At a given Fourier frequency, $f_j$, we estimate the cross-spectrum between the two light curves, by estimating the cross-periodogram\footnote{Note that an estimator of the cross-spectrum is the cross-periodogram, in analogy to the estimator of the auto-spectrum (i.e.\ PSD) which is the auto-periodogram (i.e.\ periodogram).}, $\mathscr{C}(f_j)$ \citep[e.g.][]{priestley81}, which is given by the following relation in a phasor form
\eqb
\mathscr{C}_{s,h}(f_j)=S^*(f_j)H(f_j)=\lvert S(f_j)\rvert\lvert H(f_j)\rvert e^{i\left(\phi_H(f_j)-\phi_S(f_j)\right)}
\label{eq:cross_spec}
\eqe
in which $S(f_j)$ and $H(f_j)$ are the discrete Fourier transforms of $s(t)$ and $h(t)$, respectively, with phases $\phi_S(f_j)$ and $\phi_H(f_j)$ and amplitudes $\lvert S(f_j)\rvert$ and $\lvert H(f_j)\rvert$, respectively. The asterisk denotes complex conjugation.\par
We average the complex cross-periodogram estimates, coming from all the observations, over a number of at least 10 consecutive frequency bins, yielding $m$ average cross-periodogram estimates, $\left<\mathscr{C}_{s,h}(f_{{\rm bin},i})\right>$ at $m$ new (averaged) frequency bins $f_{{\rm bin},i}$ for $i=1,2,\ldots,m$. Then, for each average cross periodogram estimate we derive its complex argument i.e.\ its angle with the positive real axis, known also as \textit{phase}, $\phi(f_{{\rm bin},i})$, 

\eqb
\phi(f_{{\rm bin},i})&=&\arg\left[\left<\mathscr{C}_{s,h}(f_{{\rm bin},i})\right>\right]\\ \nonumber
&=&\arctan\left[\frac{\im\left[\left<\mathscr{C}_{s,h}(f_{{\rm bin},i})\right>\right]}{\re\left[\left<\mathscr{C}_{s,h}(f_{{\rm bin},i})\right>\right]}\right]
\label{eq:phase}
\eqe
in which and $\re$ and $\im$  denote the real and imaginary parts of the complex quantity, respectively. Finally, we convert the estimated phase to physical time units
\eqb
\tau(f_{{\rm bin},i})=\frac{\phi(f_{{\rm bin},i})}{2\pi f_{{\rm bin},i}}
\label{eq:tl_data}
\eqe
For each time-lag estimate we calculate the corresponding standard deviation, $\operatorname{std}\{\tau(f_{{\rm bin},i})\}$ via equations 16 and 17 in \citet{nowak99}.\par
At the same time from the cross-periodogram we estimate the coherence between $s(t)$ and $h(t)$ as a function of Fourier frequency \citep{vaughan97}
\eqb
\gamma^2_{s,h}(f_j)=\frac{\lvert\langle \mathscr{C}_{s,h}(f_j)\rangle\rvert^2}{\langle\lvert S(f_j)\rvert^2\rangle\langle\lvert H(f_j)\rvert^2\rangle}
\label{eq:coher}
\eqe
For each coherence estimate we calculate the corresponding standard deviation via equation 8 in \citet{nowak99}. The coherence takes values between 0 and 1 and it is a measure of the linear correlation between the two light curves at a given Fourier frequency. A very important cautionary point is that small coherence values correspond to uncorrelated phases, $\phi_S(f_j)$ and $\phi_H(f_j)$, whose differences are actually depicted by the $\phi(f_{{\rm bin},i})$ (averaged over a range of frequencies). Thus, for uncorrelated phases, $\phi(f_{{\rm bin},i})$ has a rather uniform distribution in the range $(-\pi,\pi]$ (due to phase-wrapping) that averages always to zero. That means that for small coherence values we get a time-lag of 0 that has small uncertainties, due to the large number of averaging points, appearing statistical meaningful even if there is not a real correlation between the phases and hence no meaningful time delay. The binning scheme that we have chosen is the following: $(7.65\times10^{-6},4\times10^{-5}),(4\times10^{-5},8\times10^{-5}),(8\times10^{-5},2\times10^{-4}),(2\times10^{-4},4\times10^{-4}),(4\times10^{-4},8\times10^{-4}),(8\times10^{-4},1.5\times10^{-3}),(1.5\times10^{-3},3\times10^{-3}),(3\times10^{-3},6\times10^{-3}),(6\times10^{-3},1.2\times10^{-2})$ and $(1.2\times10^{-2}, 0.025)$. The corresponding number of estimates included in each bin, before the averaging procedure is: 13, 15, 46, 76, 151, 267, 569, 1138, 2278 and 4930, respectively, which is adequate for any sort of statistical analysis involving the derivation of mean values and standard deviations.\par
In Fig.~\ref{fig:tlSpec} we show the time-lag spectrum of \n7 we use the combined EPIC light curves between the soft ($0.5-1.5$ keV) and the hard ($2-4$ keV) X-ray energy bands, coming from all the observations. The filled circles correspond to the time-lag estimates with a coherence greater than 0.2 and the open circles to those time-lag estimates with coherence smaller than 0.2. In a recent paper, \citet{epitropakis16} present results regarding the statistical properties of Fourier based time-lag estimates with respect to their coherence. They find that, in the case in which the light curves, used for the time-lag estimation, have an average signal to noise ratio equal to 3, and the final time-lag estimates are based on the averaging of at least 10 cross-periodograms, the measured time-lags are reliable estimates of the intrinsic time-lags spectrum as long as the coherence is greater than $0.1-0.2$ (see their fig.~D13). Although our time-lags estimation is not exactly identical to \citet{epitropakis16} 
recipes, the number of cross-periodograms that we average at high frequencies is much higher than 10. Therefore, we believe that out time lags spectrum does provide a reliable estimate of the intrinsic time-lags spectrum at all frequencies below the limit where the coherence becomes equal or greater to 0.2. Note, that the time-lag spectrum has also been estimated with the \citet{epitropakis16} method and the results are identical.

\subsection{Absence of ks hard time delays at low frequencies}
\label{ssect:noksTL}
A very interesting point is that our time-lag estimates are of the order of few hundred seconds. \citet{zoghbi13} (ZO13, hearafter) presented the time-lag spectrum of \n7 between $4-5$ and $6-7$ keV energy bands, using the shortest \textit{XMM-Newton} observations obtained during 2006 (obs IDs: 0311190101, Sect.~\ref{sect:obs_datRed})\footnote{\label{noObsZO13}For the time-lag estimation in ZO13 the authors write that they used both observations obtained during 2001 and 2006 (obs IDs: 0111790101 and 0311190101, respectively). After a private communication with Dr.~Abdu Zoghbi it turned up that they only used the 0311190101.}. They found that the $6-7$ keV energy band, where the \fa line peaks, lags behind the $4-5$ keV energy band by a positive time-lag spectrum of a power-law form whose time-lag estimates from $10^{-5}$ to $10^{-4}$ Hz are of the order of $2-5$ ks (figure~6, top panel in ZO13). For the time-lag estimation ZO13 used the following binning scheme before averaging i.e.\ $(10^{-5},3\times10^{-5})
,(3\times10^{-5},6\times10^{-5}),(6\times10^{-5},10^{-4}),(10^{-4},2\times10^{-4}),(2\times10^{-4},5\times10^{-4})$ and $(5\times10^{-4},10^{-3})$ Hz. Thus, the number of points entering each frequency bin is 2, 2, 3, 8, 22 and 34, respectively, which is extremely small for the lowest three frequency bins (low frequencies) for any sort of statistical analysis.\par
\begin{figure}
\includegraphics[width=3.3in]{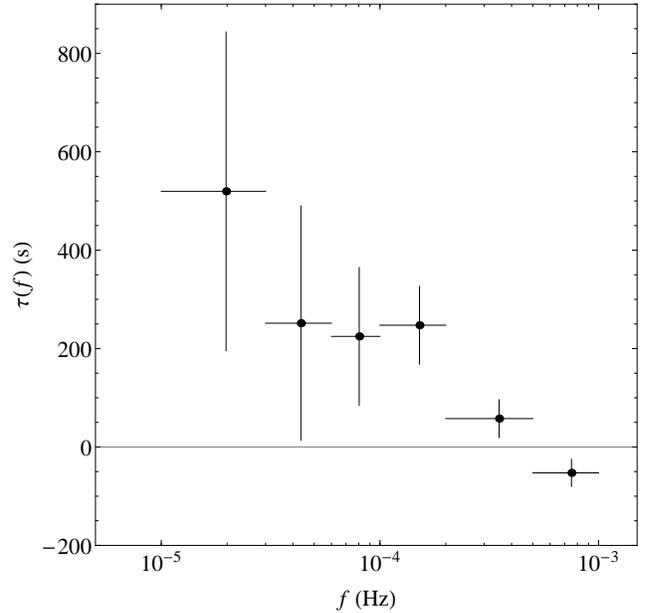}
\caption{Time-lag spectral estimates for \n7 between $4-5$ and $6-7$ keV energy bands.}
\label{fig:tlIron}
\end{figure}
\begin{figure*}
\hspace*{-31.22em}\includegraphics[width=3.3in]{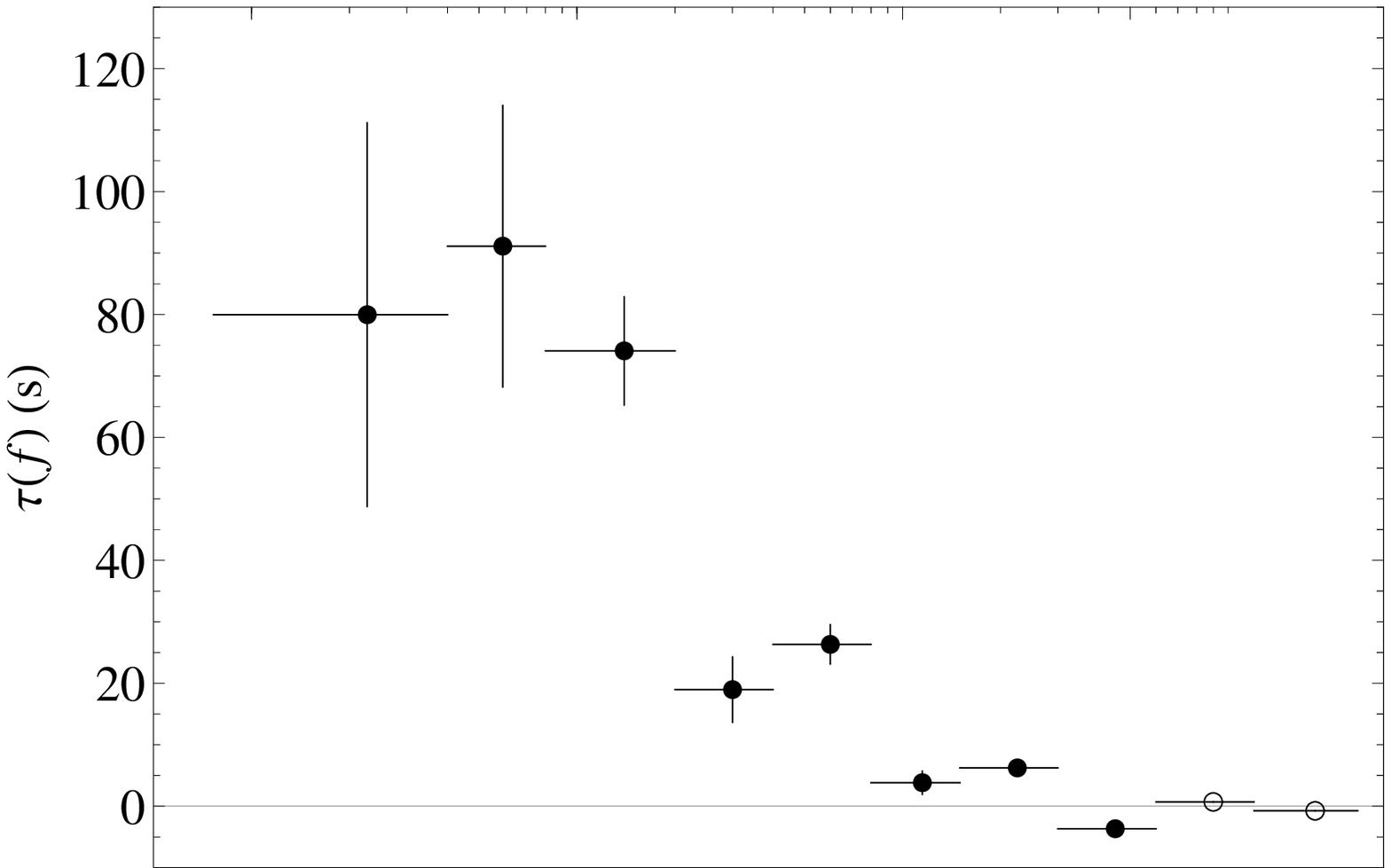}\\[-1.65em]
\parbox{1\linewidth}{\hspace{0.1em}\includegraphics[width=3.303in]{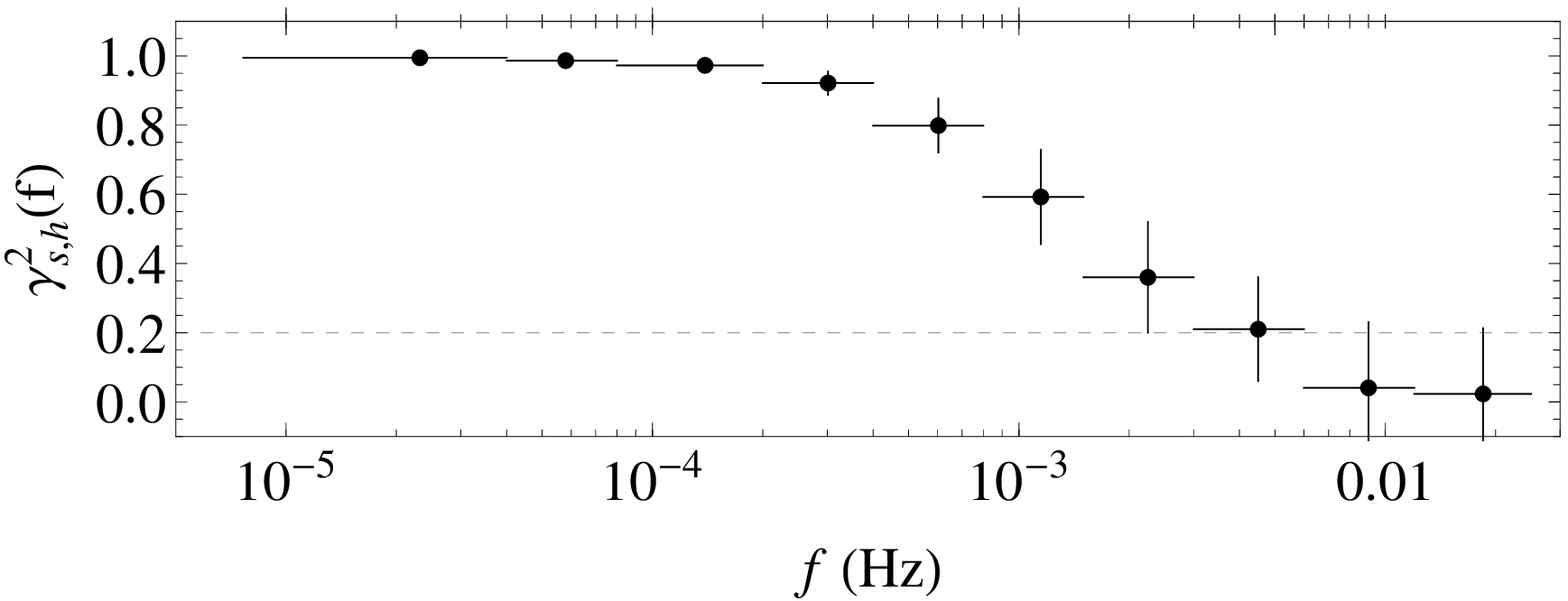}}\\[-29.35em]
\hspace{28em}\includegraphics[width=3.38in]{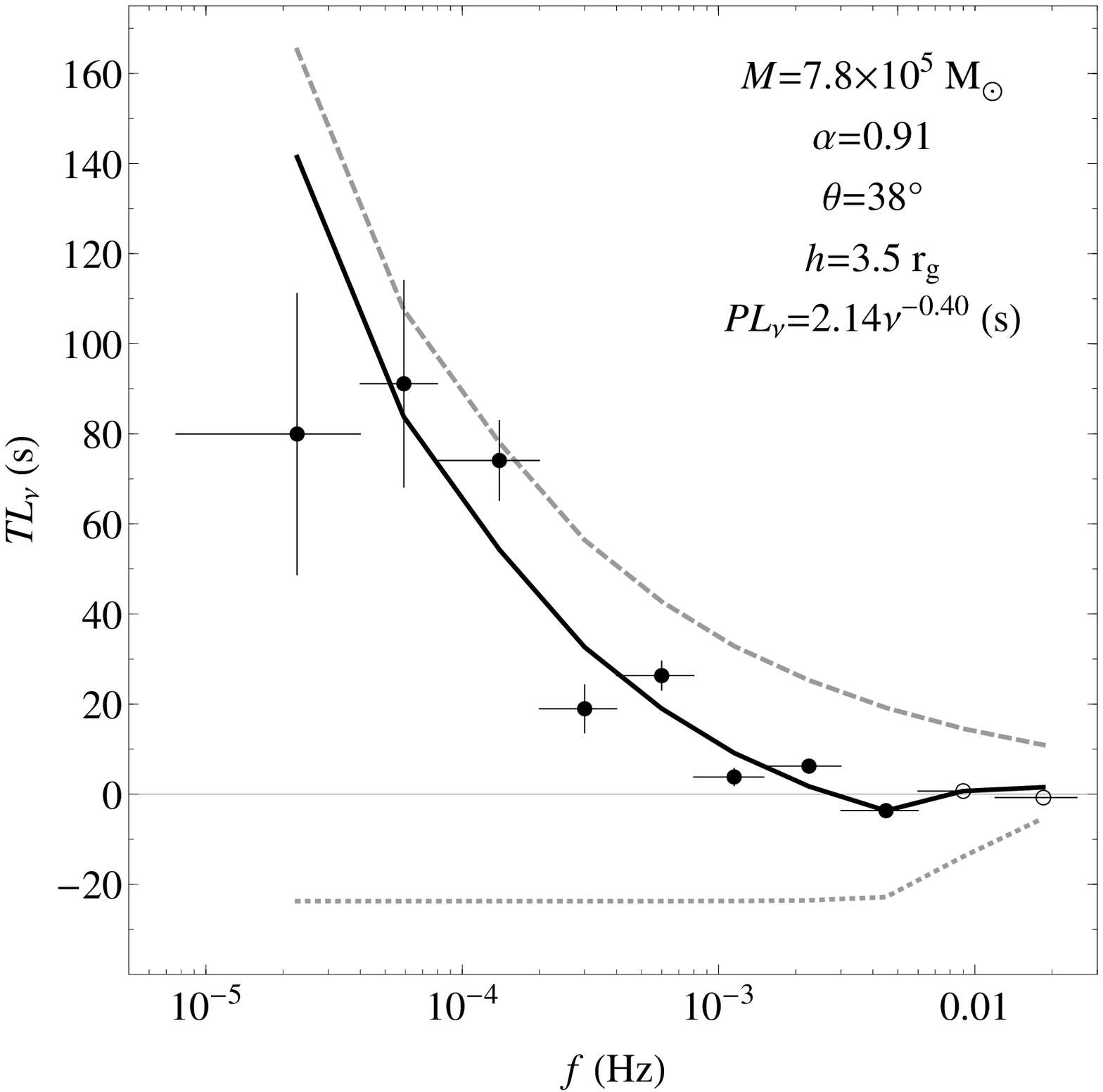}
\caption{Time-lag spectral estimates and best-fitting GR modelling for \n7 between $0.5-1.5$ and $2-4$ keV X-ray energy bands. Left-hand panel: The time-lag spectrum is shown in the upper panel and the the coherence estimates, for the same frequency bins, are plotted the lower panel. The filled/opened circles correspond to time-lag estimates with coherence greater/smaller than 0.2, respectively. Right-hand panel: The overall best-fitting GR model (solid black line) consisting of the negative GR reflection (dotted grey line) component and the positive power-law component (dotted grey line). Note that the lack of features in the GR component is due to the binning of the model as described in \citet{emmanoulopoulos14}.}
\label{fig:tlSpec}
\end{figure*}

We repeat the time-lag estimation using all four observations, in the same energy bands employing exactly the same binning scheme as ZO13. The resulting time-lag spectrum is shown in Fig.~\ref{fig:tlIron} and as we can see the corresponding estimates have the same order of magnitude as our previous time-lag estimates, between $0.5-1.5$ and $2-4$ keV energy bands (Fig.~\ref{fig:tlSpec}), but they have larger errors due to the lower-signal to noise. In this case the number of estimates, within each frequency bin, is 7, 11, 17, 37, 114 and 190, making the statistical analysis, for the estimation of the mean and the standard deviation, much more robust.\par
The appearance of ks time-delays in ZO13 is caused due to the fact that the authors tried to extend the the time-lag estimates down to $10^{-5}$ Hz using a single observation lasting around 80 ks, ending up with very few estimates at low frequencies. In Appendix~\ref{app2:timeLagIncosist} we reproduce the results of ZO13 proving in this way that the binning, rather than the actual time-lag estimation method, is the actual problem. Note that \citet{epitropakis16} have shown that reliable time-lags estimation is only possible when one averages many light curve segments (at least more than $10-20$) during the estimation of the average cross-spectrum (and consequently time-lag spectrum).\par

\subsection{Modelling of the time-lag spectra}
\label{ssect:TL_data_model}
In order to model the time-lag spectra of \n7, we employ the fully general relativistic (GR) modelling method presented in \citet{emmanoulopoulos14} for the `lamp-post geometry'. Thus, since in the Fourier domain the general relativistic impulse response functions (GRIRFs) for the system depend on the black hole (BH) mass, $M$, BH spin parameter, $\alpha$, viewing angle, $\theta$, and height, $h$, of the X-ray source above the disc, the same applies to the time-lag spectra, $\tau_f(M,\alpha,\theta,h)$, in the time domain. Moreover, we add a second component, of a power-law form, $A \nu^{-s}$, to parametrize the time-lag spectra at low Fourier frequencies (typically lower than $10^{-4}$ Hz), where the time delays are positive. Thus the overall model can be written in the following form

\eqb
TL_f(\mathbf{v})=\tau_f(M,\alpha,\theta,h)+PL_\nu(A,s)
\label{eq:overall_tlSpec}
\eqe
with $\mathbf{v}$ being the five-dimensional model parameter vector, $\mathbf{v}=\{M,\alpha,\theta,h,A,s\}$. Note that during the modelling, we select only the physical meaningful time-lag estimates with coherency values greater than 0.2 (filled circles).

As it is described in \citet{emmanoulopoulos14} another free parameter of the system is the reflection fraction between the continuum and the reflected components, at the soft energy band, which is parametrized as $f$ \citep[equation~3 in][]{emmanoulopoulos14}. However, including $f$ in the fitting procedure would result in a prohibitively large number of free model parameters in comparison to the limited number of fitted time-lag estimates (i.e.\ eight data points). Thus, as it is described in \citet{emmanoulopoulos14} we freeze its value to 0.3 following the results of \citep{crummy06}. Note that some of the model parameters i.e.\ $\alpha$, $\theta$ and $f$, can be independently constrained by fitting to the X-ray energy spectrum with reflection models. This sort of analysis will be performed in a forthcoming paper which will be entirely dedicated to the X-ray spectral properties of the source. 

The best-fitting model parameters are the following (Fig.~\ref{fig:tlSpec}, right-hand panel): for the negative GR component (dotted grey line) $h=3.5^{+2.3}_{-1.3}$ \rg, $\theta=38^{+12.3}_{-11.7}$\degr, $M=7.8^{+6.9}_{-5.8}\times10^5$ \ms, $\alpha=0.91^{\text{---}}_{-0.41}$, and for the power-law component (dashed grey line) $A=2.14^{+0.17}_{-0.12}$ s and $s=0.40^{+0.09}_{-0.07}$. These best-fitting model parameters yield a $\chi^2$ value of 45.54 for 2 d.o.f. corresponding to a practically zero $H_0$ probability of $1.29\times10^{-10}$. However, the inclusion of the negative GR component seems to be mandatory, since a simple power-law model model is consistent with zero (due to the very high statistical weight of the negative point), $A=0.00^{+0.11}_{-0.13}\times10^{-3}$ s and a slope of $s=1.74^{+0.16}_{-0.12}$ ($\chi^2$ value of $1.90\times10^6$ for 6 d.o.f).\par
The large $\chi^2$ value that we get even after including the GR component is a result of the model not capturing the frequency dependence of the positive lags i.e.\ $10^{-5}-10^{-3}$ Hz. The only negative point with physical meaning is the 8\textsuperscript{th} point at 4.5 mHz with a value of -3.66 s and an error of 2.90 ms, making it a very significant detection of negative reverberation. This point is around 1266 standard deviations away from zero thus the probability of getting a positive value from a normal distribution having a mean value and a standard deviation of -3.66 s and 2.90 ms is practically zero (i.e.\ $2.82\times10^{-345880}$).\par
\begin{figure}
\psfrag{x}{\hspace{-3em}$\re\left[\left<\mathscr{C}_{s,h}(f_{{\rm bin},i})\right>\right]$}
\psfrag{y}{\hspace{-3em}$\im\left[\left<\mathscr{C}_{s,h}(f_{{\rm bin},i})\right>\right]$}
\includegraphics[width=3.3in]{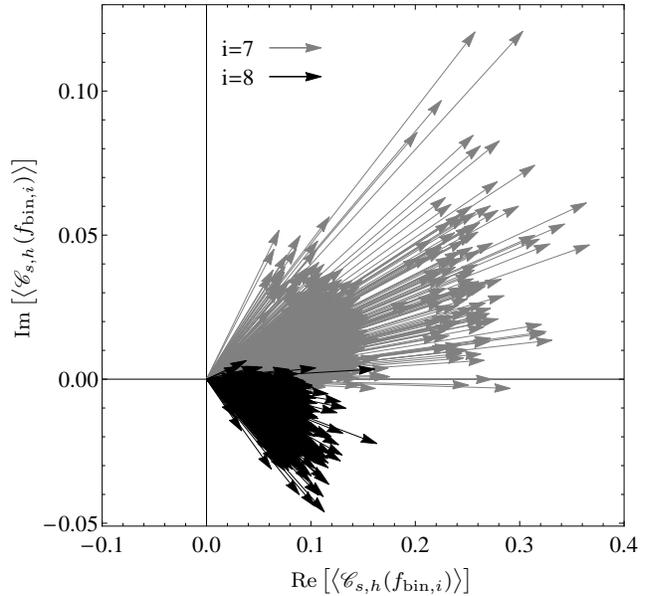}
\caption{The distribution of the average cross-spectra estimates for the seventh ($i=7$, grey arrows) and eighth ($i=8$, black arrows), frequency bins respectively. The arrows depict 1000 random draws from the distribution of the average cross-spectra estimates for each frequency bin.}
\label{fig:mcVecs}
\end{figure}

In order to visualize how significantly different is the eighth negative measurement with respect to the others we will compare it with the seventh positive measurement which is equal to $6.23\pm1.12$ s using the number of estimates that we used to compute each one individually. As we discussed in Sect.~\ref{ssect:TL_data_estim} the seventh and the eighth frequency bin consist of 569 and 1138 cross-spectra estimates, respectively. Thus, based on these estimates we produced, for each frequency bin, a two-dimensional empirical distribution of the average cross-spectrum which describes the real and imaginary parts of it, and we derive 1000 random variates. The results of the random draws are shown in Fig.~\ref{fig:mcVecs} in the form of vectors in the complex plane, with grey and black colours corresponding to the vectors of the seventh and eighth frequency bin, respectively. It is clear that the two population of vectors are genuinely different, and indeed a Kolmogorov-Smirnov test yields that the null hypothesis that the average cross-spectra estimates in the eighth frequency bin (yielding the negative time-lag estimate) are distributed according to the estimates of the seventh bin (yielding the positive time-lag estimate) is rejected at the 5 percent significance level with a practically zero probability (around $10^{-688}$). Thus, this is a genuine negative time-lag estimate originating from a significantly different parent distribution than the adjacent lower frequency positive time-lag estimate. Note that by performing a similar comparison between the negative point and any other positive time-lag estimate we will get an even more significant deviation as all the positive time-lag values are larger than 6.23 s (seventh point).\par
The single negative point at 4.5 mHz is the one that tunes the best-fitting GR component giving greater emphasis in the height of the X-ray source. Thus, the estimated uncertainties for the best-fitting parameters $\theta$, $M$ and $\alpha$ are rather large with the only exception those of the height, $h$. As it is shown in \citep{wilkins13,cackett14,emmanoulopoulos14} the height parameter is the one that gives the most prominent peaks and wiggles in the time-lag spectra, and thus this is the one that will be driven firstly from any negative point. From figures B1, B2 and B3 in the appendix B of \citet{emmanoulopoulos14} we can see that the height is the dominant fitting parameters as its fluctuations cause great differences in the resulting time-lag spectra.\par 
Moreover the level of the GR component is actually tuned by the low frequency part of the time-lag spectrum. Undoubtedly, the positive power-law model is the dominant component for the overall time-lag spectrum and thus its normalization fixes automatically the level of the GR component. Interestingly, from figures B1, B2 and B3 in \citet{emmanoulopoulos14} we see that only the height parameter shows great variations at the low frequency end of the time-lag spectrum, and thus by fixing the power-law normalization automatically the height of the X-ray source is further constrained. Note, that there are no degeneracies between the power-law and the GR component but there are degeneracies between the various parameters within the GR component that are discussed in detail in \citet{emmanoulopoulos14}. 

\subsection{Energy resolved time-lag spectrum}
\label{ssect:TL_energyResolved}
\begin{figure}
\includegraphics[width=3.3in]{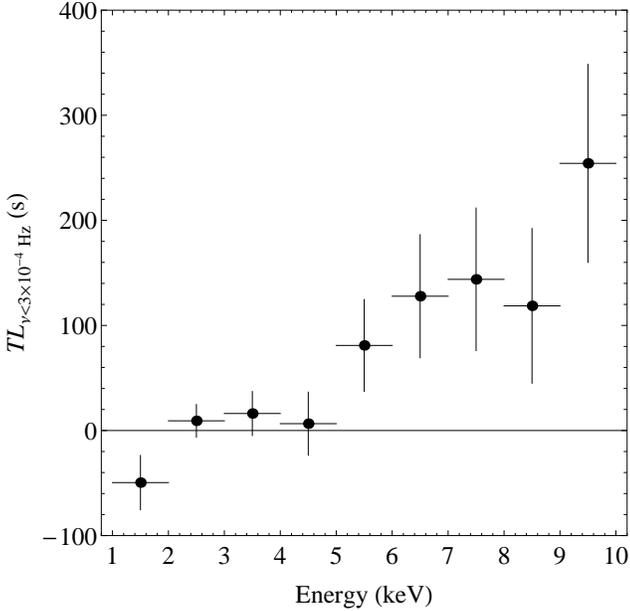}
\caption{The energy resolved time-lag spectrum of \n7. We consider all the frequencies below $3\times10^{-4}$ Hz and we use as a reference light curve, that in the $1-10$ keV energy range.}
\label{fig:energyResolvedtlSpec}
\end{figure}

In this section we estimate the energy resolved time-lag spectrum of \n7. We calculate the time-lag spectra between light curves in nine different energy bands (i.e.\ $1-2,2-3,\ldots,9-10$ keV) and a reference energy-band light curve covering the whole $1-10$ keV, for which we exclude each time the current energy band (i.e.\ $2-10,1-2\cup3-10,\ldots,1-9$ keV). The time-lag spectrum between a given light curve and the reference energy-band light curve is done exactly as described in Section~\ref{ssect:TL_data_estim}. Then, for each energy band we average all the cross-periodogram estimates (equation~\ref{eq:cross_spec}) for all the frequencies below $3\times10^{-4}$ Hz and we derive a mean time-lag value using equation~\ref{eq:tl_data}. The same exactly procedure is done for the coherence and the error estimates. The resulting energy-dependent time-lag spectrum is shown in Fig.~\ref{fig:energyResolvedtlSpec}, showing a clear constant increase between 4 and 10 keV.\par
As in Sect.~\ref{ssect:noksTL}, our energy resolved time-lag spectral estimates are much smaller from those presented in figure~7 in ZO13 which are of the order of few ks. Similarly as before, this discrepancy is attributed to the very small number of points, entering the bins below $3\times10^{-4}$ Hz, leading to an overestimation of the resulting time-lag spectrum across the entire energy range.

\section{SUMMARY AND DISCUSSION}
\label{sect:summ_disc}
We have performed a thorough X-ray variability analysis of all four \textit{XMM-Newton} observations of the low BH mass AGN \n7. These observations consist of two archival data sets of short duration, 45 and 84 ks, obtained during 2001 and 2006, respectively, and two newly obtained long observations of 130 and 140 ks, during 2013. We have performed an energy resolved PSD analysis, using a new \textit{state-of-the-art} analytical method (which will be described in a forthcoming paper) that takes into account all the spectral pathologies induced by the window function. For our \textit{XMM-Newton} data, only a very marginal contamination, from red-noise leak, is affecting the 2001 data and it is taken into consideration. Finally, we perform a classical X-ray time-lag analysis to probe possible reflection signatures from the close BH environment. The main results from our analysis are summarised below.
\begin{itemize}
\item The best-fitting double bending PSD model, in the $0.5-10$ keV energy band, has a high-frequency PSD bend, $f_{\rm h}$ of $6.71\times10^{-5}$ Hz, a medium-frequency PSD index, $\alpha_{\rm m}$, of 0.51 and a high-frequency PSD index, $\alpha_{\rm h}$, of 1.99. In all our analysis the low-frequency PSD index, $\alpha_{\rm l}$, and PSD bend frequency, $f_{\rm l}$, are fixed to 0 and $10^{-8}$ Hz, respectively.
\item We found that as the energy increases $\alpha_{\rm h}$ becomes flatter and the PSD normalization, $A$, becomes larger.
\item There is a small hint that as the energy increases $\alpha_{\rm m}$ becomes steeper and $f_{\rm h}$ is shifted towards smaller frequencies.
\item The addition of our PSD bend time-scale estimate of \n7 to the sample of GV12 yields that the bend time-scale depends linearly on mass. This conclusion is consistent with the results of GV12 under the assumptions that the medium frequency PSD slope, whose value can affect the measured PSD bend frequency, is fixed at -1 in all AGN, and both the values of BH mass and $L_{\rm bol}$ estimates are fixed to the uncertain literature values.
\item The X-ray time-lag spectra between $0.5-1.5$ and $2-4$ keV, show very weak signatures of negative time delays, and the energy-resolved time-lag spectra, do not show any indication of X-ray reverberation, as it is expected at frequencies below $3\times10^{-4}$ Hz.
\item At low frequencies our improved data show hard time-lags of few hundred seconds rather than the few ks claimed by \citet{zoghbi13}.
\end{itemize}

\subsection{The PSD bend frequency}
\label{ssect:psdBendFreq}
\citet{mchardy04}  found that, if they fixed the PSD slope below the bend frequency to be the same at all energies, the normalization increases with decreasing energy. However, as the slope of the PSD above the bend frequency is flatter at higher energies, the high energy PSD lies above the low frequency PSD at the highest frequencies sampled. Unlike in the present work, \citet{mchardy04} having only RXTE data to sample the low PSD frequencies, were unable to determine the PSD slope below the bend as a function of energy. Thus, direct comparison of normalizations is difficult. However, both \citet{mchardy04}, for the case of NGC\;4051 and the present work find that the PSD slope above the bend frequency is flatter at higher energies. This is a common behaviour seen in several AGN e.g.\ NGC\,7469 \citep[in which the PSD of even the $10-15$ keV energy band appears flatter those in the lower energy bands]{nandra01}, MCG--6-30-15 \citep{vaughan03b}, Mrk\,766 \citet{vaughan03c}.\par
Associating different variability time-scales (i.e.\ Fourier frequencies) with different locations on the accretion disc we can explain the observed PSD behaviour. In the model of propagating fluctuations \citep{lyubarskii97,kotov01} in which variations, produced in different radii in the accretion flow, propagate inwards in an optically thick corona, over the surface of the disc until it hits the X-ray emitting region, situated few $r_{\rm g}$ from the central BH, whose emission is then modulated. A geometry where high Fourier frequencies (small variations) and high X-ray energies are produced from the inner parts of the X-ray emitting regions, provides an immediate explanation of the flattening of the high, relative to the low, energy PSD at high frequencies, depicted by $\alpha_{\rm h}$. In this scenario, the steeper values of the lower energy medium frequency slope, $a_{\rm m}$, compensate for the larger normalization at higher energies, and thus the values of $F_{\rm var}$ at lower energies (Sect.~\ref{sect:fvar}), are larger from those in higher energies.\par
Currently, there is a debate whether the high-frequency bend time-scales, $f_{\rm h}$, observed in the PSDs of AGN scale linearly only with the BH mass or with both the BH mass, $M$ and the bolometric luminosity, $L_{\rm bol}$. \citet{mchardy06} using a sample of 10 AGN and 2 galactic BHs (Cyg\,X-1 and GRS\,1915+105) found that $f_{\rm h}$ scales with both $M$ and $L_{\rm bol}$ indicating that AGN are scaled-up versions of galactic BHs. In contrast, GV12 used a sample of 22 AGN and found only a very week dependence between $M$ and $L_{\rm bol}$, which becomes slightly stronger if one considers different BH masses for 2 sources (NGC\,4395 --lowest BH mass in the sample-- and NGC\,5506) excluding at the same time NGC\,6860. Our PSD bend time-scale of \n7 supports this lack of correlation between $M$ and $L_{\rm bol}$, but we have to note that this is the only source in our sample with long enough data for which both PSD slopes, $\alpha_{\rm m}$ and $\alpha_{\rm h}$ are clearly resolved, and thus $f_{\rm h}$ 
can be estimated robustly. The short duration of the \textit{XMM}-\textit{Newton} observations required GV12 to fix the medium frequency PSD slope (referred to as low-frequency PSD slope in GV12) at 1, and we do the same here as we use their PSD bend time-scales for all of the sources in their sample except for \n7. The measured bend time-scale depends on the low frequency slope and there are several examples in which slightly different values of low frequency PSD slope have been measured e.g.\ this work for \n7 in which $\alpha_{\rm m}=0.51$, MCG--6-30-15 in which $\alpha_{\rm m}=0.8$ \citep{mchardy05}. Moreover, uncertainties in the BH mass values can also affect the best-fitting results particularly in small samples \citep[e.g.\ NGC\,5506;][]{mchardy13}. Thus, this lack of correlation between $f_{\rm h}$ and $L_{\rm bol}$ can be caused due to uncertainties on the BH masses and/or $L_{\rm bol}$.\par
A very interesting point from our analysis, is that the medium-frequency spectral index, $\alpha_{\rm m}$, is around 0.5. In almost all the cases, where the high-frequency bend is very marginally or not at all resolved, it is assumed that $\alpha_{\rm m}$ is equal to 1 \citep[e.g.][]{gonzalez12} based on long-term observations \citep[e.g.][]{uttley02,markowitz03,mchardy06}. This difference, of a factor of two, affects both the values of $f_{\rm h}$ and $\alpha_{\rm h}$. The average best-fitting value of $f_{\rm h}$ and $\alpha_{\rm h}$ in the $0.5-10$ keV energy band for a fixed value of $\alpha_{\rm m}$ equal to 1, (Table~\ref{tab:psd_0.5_10_am}, first three lines i.e.\ ignoring the fourth observation) is $2.05\times10^{-5}$ Hz and 1.78, respectively. By comparing these values with the best-fitting values, with the joint best-fitting results from the three observations (Table~\ref{tab:joint_psd_energy_bands}, first line), we find a very significant difference of an order of magnitude for $f_{\rm h}$ and a 
milder one for $\alpha_{\rm h}$ of around 10 per cent. Thus, being able to determine precisely the medium-frequency PSD slope is of great importance if one wants to derive the position of high-frequency PSD bend. However, note that this rather flat value of $\alpha_{\rm m}=0.5$, could result if this part of the PSD depicts the peak of a Lorenzian component as it has been resolved in the case of Ark\,564 \citep{mchardy07}, using much longer data sets obtained by \textit{RXTE} and \textit{ASCA}.\par
Within the propagating fluctuation model, a medium-frequency PSD index of the order of 0.5 can be the result of radial damping of the fluctuations as the propagate through the accretion flow \citep{arevalo06}. The effect of damping is to reduce the amplitude of the fluctuations as they propagate inwards so that the further they go the smaller they get. By extrapolating the values shown in table 2 of \citet{arevalo06}, for a high emissivity index of $\gamma=5-6$ and a damping coefficient of the order of 2 the medium-frequency PSD index can be of the order of 0.5. Note, that a similar flat value of $\alpha_{\rm m}$ ($0.8^{+0.40}_{-0.16}$, which is consistent, within the errors with our estimate) has been also observed in the case of MCG--6-30-15 \citep{mchardy05}.

\subsection{Is there any evidence for relativistic reflection?}
\label{ssect:evid_GRreflect}
The time-lag spectrum of \n7 (Fig.~\ref{fig:tlSpec}) poses weak evidence of negative X-ray reverberation time delays between the reflected component, $0.5-1.5$ keV, and the continuum $2-4$ keV. Our GR modelling yields a best-fitting BH mass of $7.8^{+6.9}_{-5.8}\times10^5$ \ms, which is very close to the literature value ($0.87\times10^6$ \ms, Section~\ref{sect:intro}), and the height of the X-ray source, $3.5^{+2.3}_{-1.3}$ \rg. \citet{demarco13} showed that both the most negative time-lag estimate, -3.66 s, and its corresponding frequency, 4.5 mHz are related linearly (in logarithmic scale) with the BH mass. Using their scaling relations the expected values of both these quantities should be around -20 s and 1 mHz, respectively which differ significantly from our observed values. In order for the latter estimates to be consistent with the \citet{demarco13} relations the BH mass of \n7 should be around $4\times10^4$ \ms which is more than an order of magnitude less than the literature value. As it is described in \citet{emmanoulopoulos14} the scaling relation of \citet{demarco13}, derived from the first negative time-lag point, holds only for sources with relatively small X-ray source heights and similar power-law components. In our case the power-law slope is very hard, 0.4, suppressing any negative peaks at lower frequencies, thus it is not surprising that such a relation does not hold. Moreover, this deviation from the scaling relation could be the result of a relatively low contribution of reflected emission to the soft X-ray band.\par
Moreover, the absence of strong negative X-ray reverberation signatures can be justified from the fact that the X-ray spectrum of the source around the $0.5-1.5$ keV is rather flat and absorbed \citep{ebrero11} with no obvious soft-excess or even signs of strong ionised soft X-ray reflection. Similarly the X-ray spectrum presented by \citet{zoghbi13} is in agreement with this picture since the component responsible for the inner reflection contributes only around the region of 6.4 keV without affecting the $0.5-1.5$ keV energy range.\par
As we described Sect.~\ref{ssect:TL_data_model}, during the time-lag modelling procedure, we fix the value of the reflection fraction to 0.3, due to the limited number of time-lag estimates (i.e.\ eight). It is very well known \citep{wilkins13,cackett14} that the contribution of the continuum component to the reflection-dominated band and vice-versa, via $f$, can reduce the apparent time lag that is measured. Indeed, in appendix~A of \citet{emmanoulopoulos14} these dependencies are shown with respect to the reflected component. In an attempt to check the effects of the reflection fraction, in our time-lag estimates, we repeat the fitting by changing the value of $f$ to 0.6 and 0.9. For $f=0.6$ the best fitting are $h=9.7^{+4.1}_{-5.4}$ \rg, $\theta=45^{+17.3}_{-19.7}$\degr, $M=10.8^{+9.2}_{-7.8}\times10^5$ \ms, $\alpha=0.81^{\text{---}}_{-0.61}$, and for $f=0.9$ the best-fitting parameters are $h=10.8^{+8.1}_{-7.4}$ \rg, $\theta=43^{+21.4}_{-22.7}$\degr, 
$M=10.5^{+10.2}_\text{---}\times10^5$ \ms, $\alpha=0.85^{\text{---}}_\text{---}$. In both cases, the values of the $\chi^2$ statistics are much larger than that of $f=0.3$, thus we have good grounds to believe a physical scenario in which the contribution of the reflection spectrum in the soft band is indeed around 30 per cent of the total observed flux.\par
Finally, the monotonic increase in the energy resolved time-lag spectral estimates, above $4$ keV (Fig.~\ref{fig:energyResolvedtlSpec}) dictates that it arises from the continuum component itself, which is modelled as the power-law in the time-lag spectrum (Fig.~\ref{fig:tlSpec}). This is a model expectation based on an extrapolation of results from X-ray binaries rather than a prediction of detailed theories of the continuum emission. Similar behaviour has been observed in several other AGN e.g.\ 1H\;0707-495, Ark\;564, Mrk\;335 \citep{kara13a,kara13c} and can be interpreted using the previously mentioned propagating fluctuations model. Fluctuations in the accretion flow, which are introduced at different radii/time-scales, propagate  inwards  on  diffusion  time-scales, causing the observed power-law shape of the lag spectrum. Thus, these very low frequency variations presumably originate from regions further from the centre, where the temperature of the disc is cooler.

\section*{Acknowledgments}
DE and IMM acknowledge the Science and Technology Facilities Council (STFC) for support under grant ST/G003084/1. SV acknowledges the STFC for support under grant ST/K001000/1. DE is grateful to Dr.~Abdu Zoghbi who provided all the necessary information about the analysis methods used in \citet{zoghbi13} during the workshop `The X-ray Spectral-Timing Revolution' in February 2016, Leiden, the Netherlands. This research has made use of NASA's Astrophysics Data System Bibliographic Services. Finally, we are grateful to the anonymous referee for the very useful comments and suggestions that helped improved the quality of the manuscript.

\appendix
\section[]{PSD model parameter confidence limits}
\label{app1:confidence_limits}

\begin{figure*}
\includegraphics[width=3.3in]{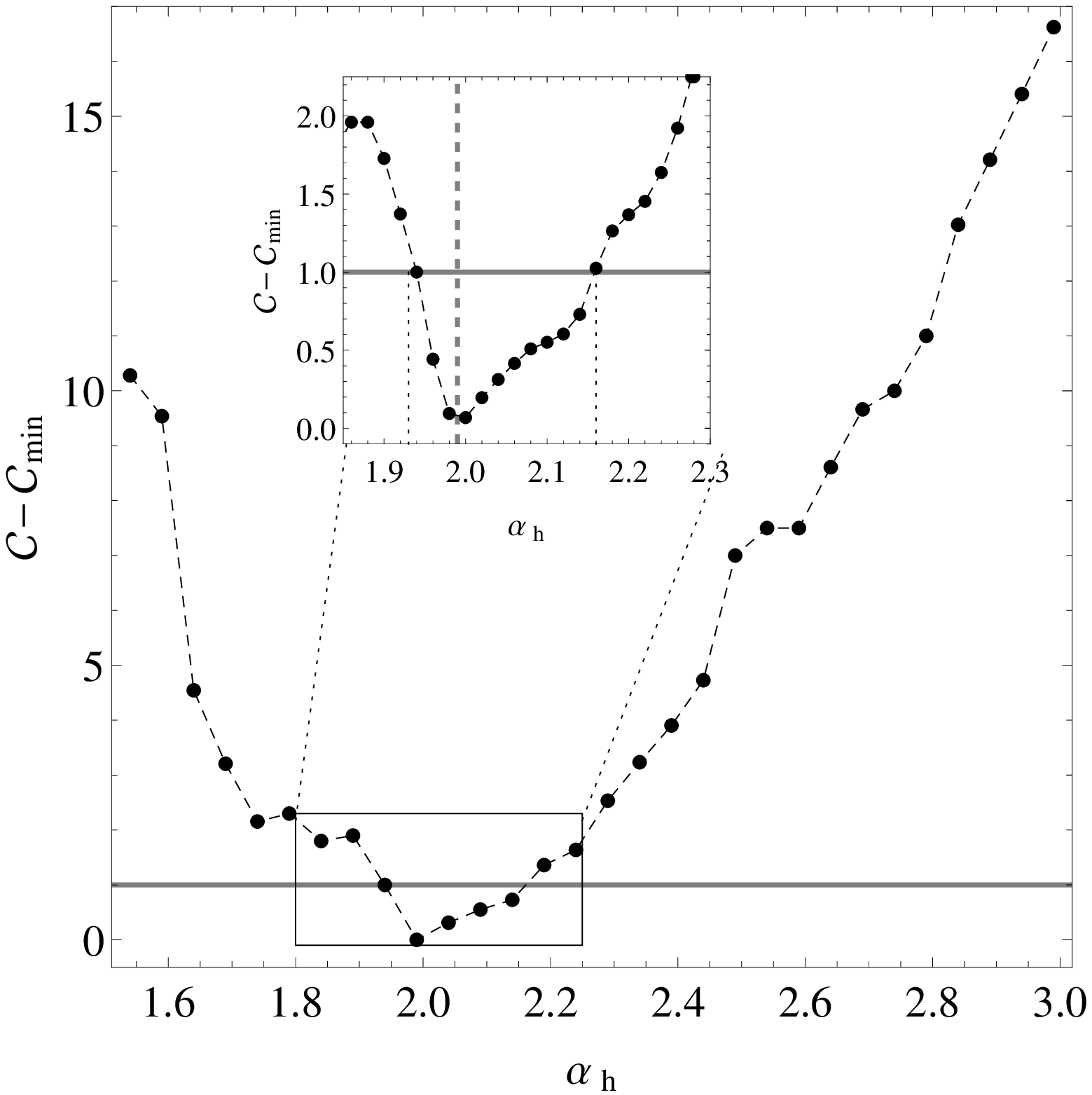}\hspace{1em}
\includegraphics[width=3.25in]{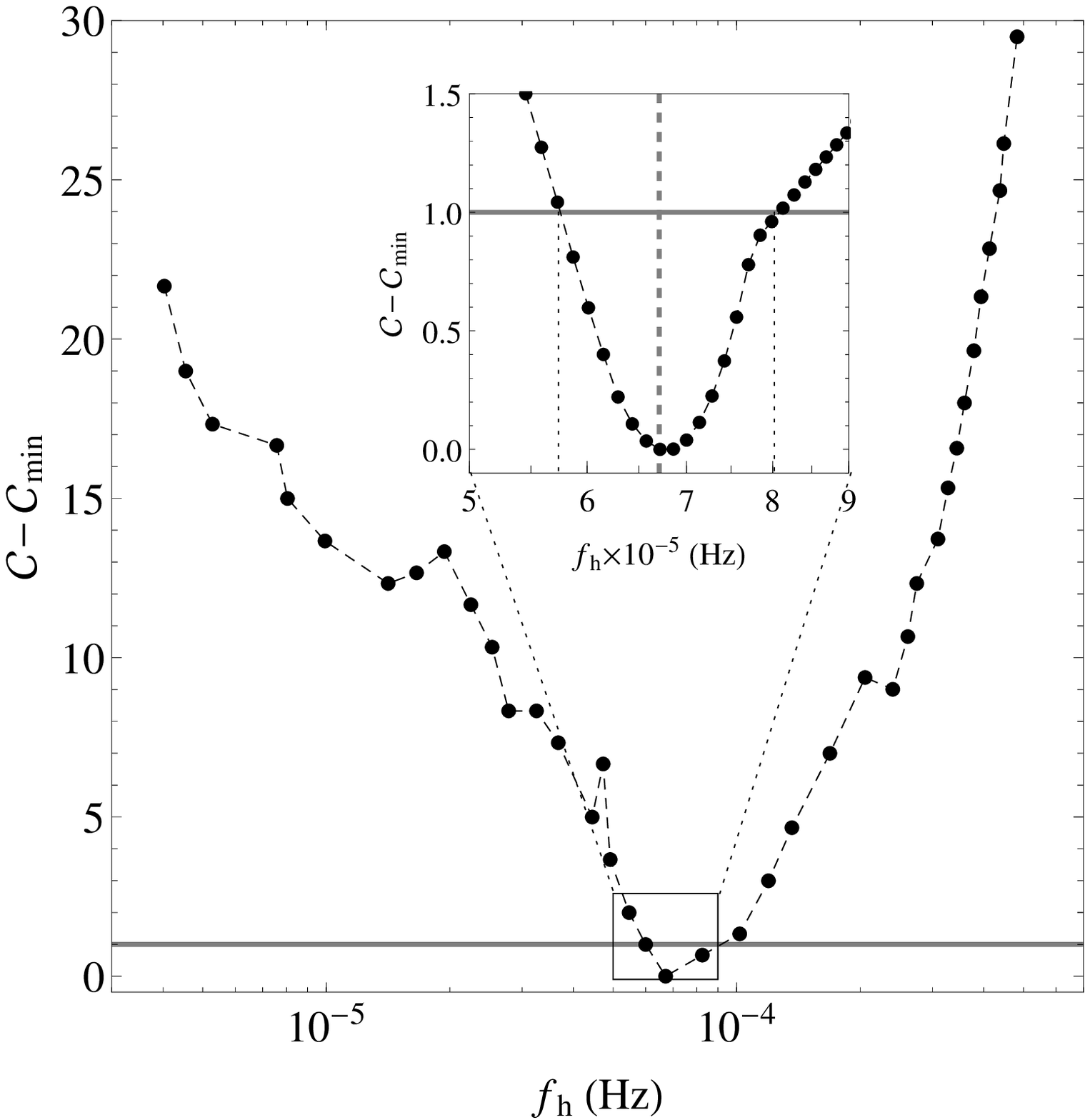}\\[1em]
\includegraphics[width=3.21in]{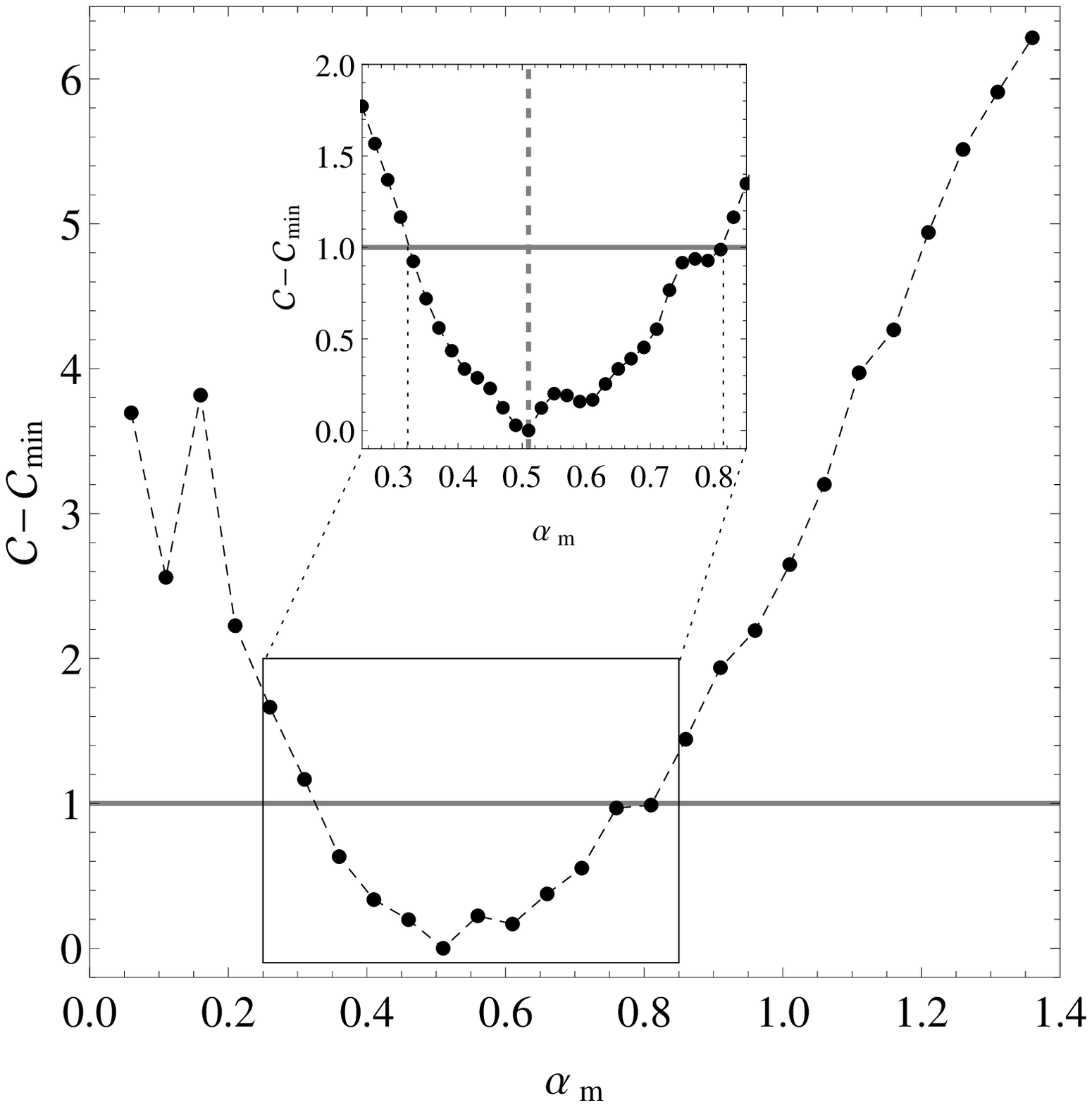}\hspace{1.2em}
\includegraphics[width=3.25in]{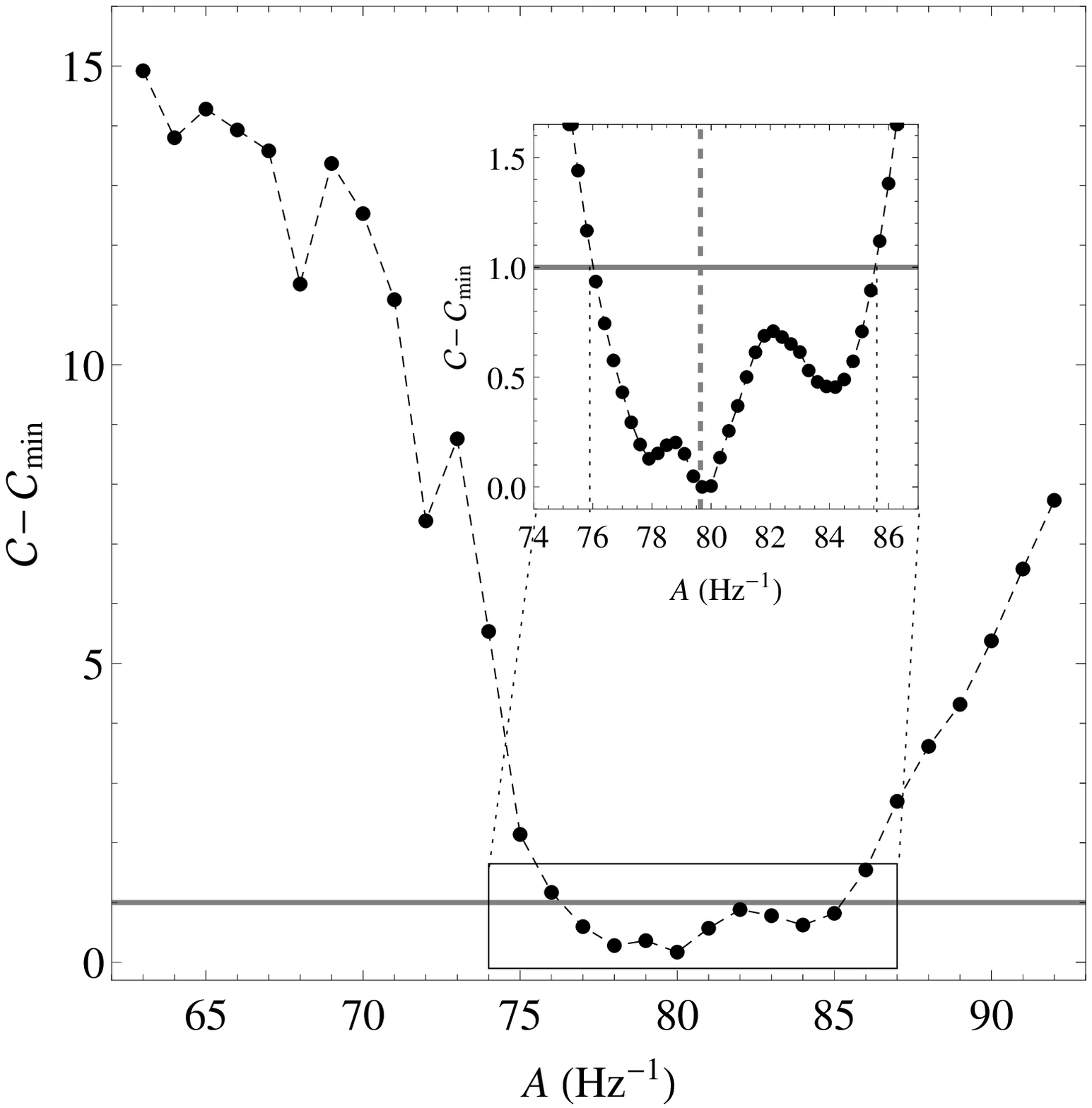}
\caption{The log-likelihood PSD model parameter profiles, $\Delta\mathcal{C}\equiv\mathcal{C}-\mathcal{C}_{\rm min}$, for the joint fits in the $0.5-10$ keV energy range. The filled circles correspond to the actual log-likelihood estimates i.e.\ grid points, and the dashed line depicts the linear interpolation estimates. The inset zooms in the region around the best-fitting model parameter (vertical dashed line), and the dotted vertical lines around it depict its 68.3 per cent single confidence limits that correspond to $\Delta\mathcal{C}=1$.}
\label{fig:singleConfContour}
\end{figure*}

In Sect.~\ref{ssect:psd0p5t10freeAm} we estimated the best-fitting PSD model for the ensemble of the observations in the 0.5-10 keV energy range. In order to visualize the procedure that we used to derive the 68.3 per cent confidence limits for the best-fitting model parameters, in Fig.~\ref{fig:singleConfContour} we show the log-likelihood difference profiles, $\Delta\mathcal{C}\equiv\mathcal{C}-\mathcal{C}_{\rm min}$, as a function of each model fitting parameter. Additionally, the inset of each panel shows with a finer grid-resolution the region around the estimated 68.3 single confidence limits that corresponds to $\Delta\mathcal{C}=1$, which is depicted by the dotted vertical lines.\par
\begin{figure*}
\hspace{-1em}
\parbox{1\linewidth}{\includegraphics[width=3.8in]{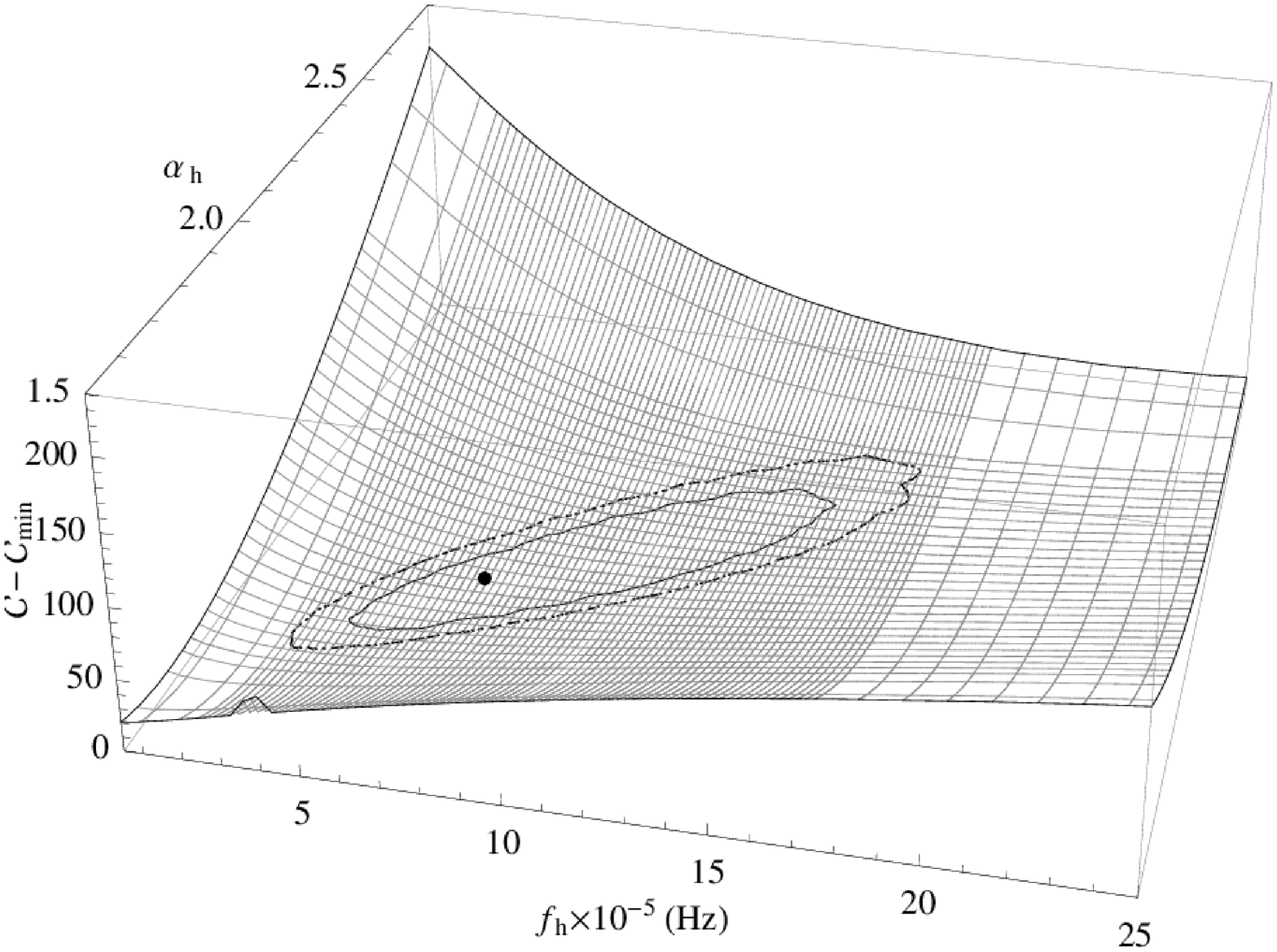}\hspace{0em}\parbox{1\linewidth}{
\vspace*{-20em}\includegraphics[width=3.2in]{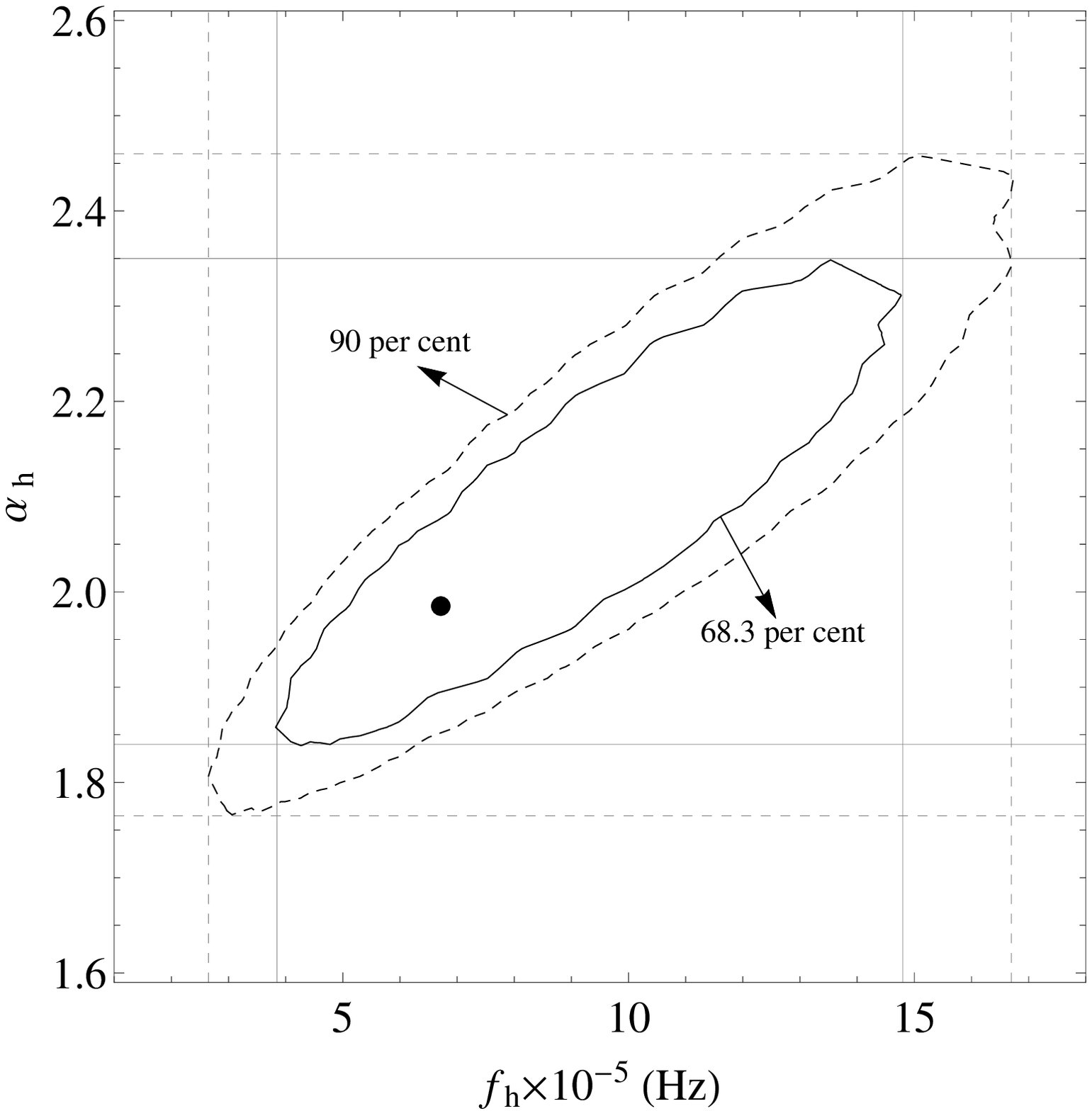}}}
\caption{The joint confidence contours of $f_{\rm h}$ and $\alpha_{\rm h}$ around their best-fitting values (black filled symbol), for the case of the $0.5-10$ keV best-fitting source PSD model. Left-hand panel: The 3-dimensional surface $\Delta\mathcal{C}$ (grey adaptive mesh) estimated at the crossing points of the grid. The solid and dashed ellipsoidal lines depict the joint 68.3 and 90 per cent joint confidence contours of $f_{\rm h}$ and $\alpha_{\rm h}$, that correspond to a $\Delta\mathcal{C}$ of 2.30 and 4.61, respectively. Right-hand panel: The 2-dimensional joint confidence contours, 68.3 (solid line) and 90 (dashed line) per cent for $f_{\rm h}$ and $\alpha_{\rm h}$.}
\label{fig:jointConfContour}
\end{figure*}

Finally, we estimate the joint confidence contours for the best-fitting PSD model parameters $f_{\rm h}$ and $\alpha_{\rm h}$ in the energy range $0.5-10$ keV. In the left panel of Fig.~\ref{fig:jointConfContour} we show the form of the $\Delta\mathcal{C}$ surface (left-hand panel, grey adaptive mesh), which is estimated at the crossing points of the grid. From this surface we can derive the 68.3 and 90 per cent joint confidence contours for $f_{\rm h}$ and $\alpha_{\rm h}$, lying at a height $\Delta\mathcal{C}$ of 2.30 (solid black line) and 4.61 (dashed black line), respectively, by estimating the tangent lines to these contours (Fig.~\ref{fig:jointConfContour}, right panel). This yields for $f_{\rm h}$ $\left[3.85,14.82\right]\times10^{-5}$ Hz and $\left[2.65,16.71\right]\times10^{-5}$ Hz and for $\alpha_{\rm h}$ $\left[1.84,2.35\right]$ and $\left[1.77,2.65\right]$, respectively.

\section[]{Time-lag estimation inconsistencies}
\label{app2:timeLagIncosist}
\begin{figure*}
\includegraphics[width=3.3in]{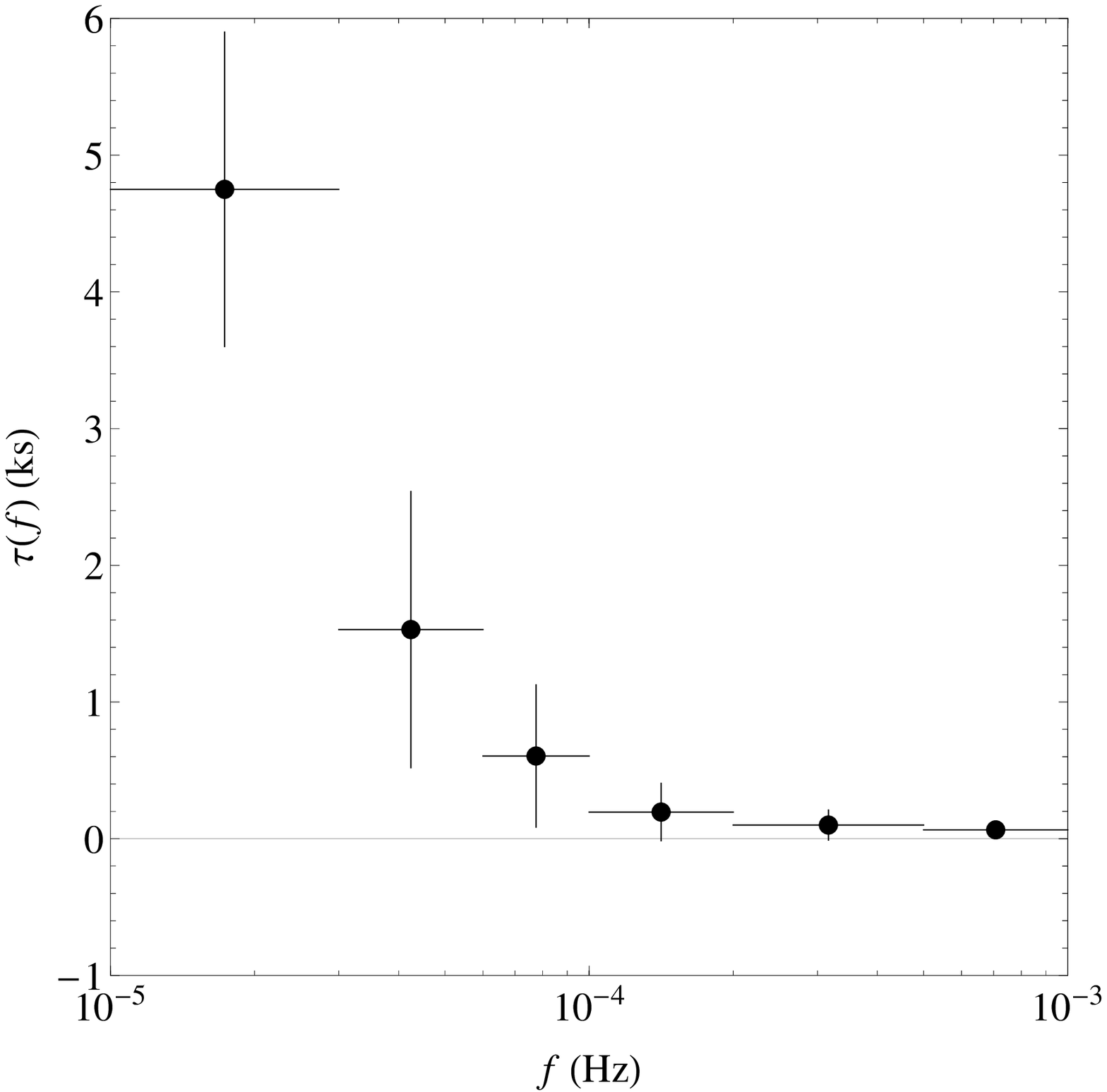}
\caption{Time-lag spectrum reproduction of ZO13 between $4-5$ and $6-7$ keV energy bands.}
\label{fig:reprodZO13}
\end{figure*}

In this section we reproduce the time-lag spectrum of \n7 as it is precisely described by ZO13. We use the single observation obs ID: 0311190101 (see footnote \ref{noObsZO13}) and the same binning scheme as described in Sect.~\ref{ssect:noksTL}. The actual time-lag estimation method is the same as described in Sect.~\ref{ssect:TL_data_estim}. The resulting time-lag spectrum is shown in Fig.~\ref{fig:reprodZO13} and it is exactly the same as the ones shown in the top panel of figure~6 in ZO13. Thus, these overestimated low-frequency time-lag values is the result of the limited number of points entering corresponding frequency bins.

\bsp
\label{lastpage}
\end{document}